\documentstyle[prc,twocolumn,aps,epsf,eqsecnum,ifthen]{revtex}

\newcommand{\beq}{\begin{equation}}
\newcommand{\eeq}{\end{equation}}
\newcommand{\bea}{\begin{eqnarray}}
\newcommand{\eea}{\end{eqnarray}}

\newcommand{\la}{\langle}       
\newcommand{\ra}{\rangle}       

\newcommand{\lda}{\langle\!\langle}       

\newcommand{\rda}{\rangle\!\rangle}

\newcommand{\tr}{{\rm tr}}
\newcommand{\bp}{{\bbox{p}}}
\newcommand{\bv}{{\bbox{v}}}
\newcommand{\bq}{{\bbox{q}}}
\newcommand{\bk}{{\bbox{k}}}
\newcommand{\bK}{{\bbox{K}}}
\newcommand{\bx}{{\bbox{x}}}
\newcommand{\tp}{{\tilde p}}
\newcommand{\tk}{{\tilde k}}

\newcommand{\half}{{\textstyle {1\over 2}}}

\newcommand{\ibid}[3]{{\it ibid.} {\bf #1}, #3 (#2)}
\newcommand{\ptp}[3]{Prog. Theor. Phys. {\bf #1}, #3 (#2)}
\newcommand{\ijmpa}[3]{Int. J. Mod. Phys. A {\bf #1}, #3 (#2)}
\newcommand{\annrev}[3]{Ann. Rev. Nucl. Part. Sci. {\bf #1}, #3 (#2)}
\newcommand{\prep}[3]{Phys. Rep. {\bf #1}, #3 (#2)}
\newcommand{\prevc}[3]{Phys. Rev. C {\bf #1}, #3 (#2)}
\newcommand{\prevd}[3]{Phys. Rev. D {\bf #1}, #3 (#2)}
\newcommand{\prevl}[3]{Phys. Rev. Lett.\ {\bf #1}, #3 (#2)}
\newcommand{\zpa}[3]{Z. Phys. A {\bf #1}, #3 (#2)}

\newcommand{\jpg}[3]{J. Phys. G {\bf #1}, #3 (#2)}
\newcommand{\epjc}[3]{Eur. Phys. J. C {\bf #1}, #3 (#2)}
\newcommand{\plb}[3]{Phys. Lett. B {\bf #1}, #3 (#2)}
\newcommand{\npa}[3]{Nucl. Phys. {\bf A#1}, #3 (#2)}

\newcommand{\hip}[3]{Heavy Ion Physics {\bf #1}, #3 (#2)}

\begin{document}

\preprint{\\CERN-TH/2000-123\\hep-ph/0006...}

\title{Multi-boson effects in Bose-Einstein interferometry and the
       multiplicity distribution}
\author{U. Heinz$^{a,}$\thanks{On leave of absence from 
        Institut f\"ur Theoretische Physik, Universit\"at Regensburg. 
        E-mail: ulrich.heinz@cern.ch},
        P. Scotto$^b$,
        and Q.H. Zhang$^{c,d}$}
\address{$^a$Theoretical Physics Division, CERN, CH-1211 Geneva 23, 
         Switzerland\\
         $^b$Institut f\"ur Theoretische Physik, Universit\"at Regensburg,
         D-93040 Regensburg,Germany\\
         $^c$Institute of Theoretical Science and Department of Physics,
         University of Oregon, Eugene, OR 97403-5203, USA\\
         $^d$Physics Department, McGill University, Montreal, 
         Quebec H3A\,2T8, Canada}
\date{\today}

\maketitle

\begin{abstract}
Multi-boson symmetrization effects on two-particle Bose-Einstein
interferometry are studied for ensembles with arbitrary multiplicity
distributions. This generalizes the previously studied case of a
Poissonian input multiplicity distribution. In the general case we
find interesting residual correlations which require a modified
framework for extracting information on the source geometry from
two-particle correlation measurements. In sources with high
phase-space densities, multi-boson effects modify the Hanbury 
Brown--Twiss (HBT) radius parameters and simultaneously generate strong
residual correlations. We clarify their effect on the correlation
strength (intercept parameter) and thus explain a variety of
previously reported puzzling multi-boson symmetrization phenomena. 
Using a class of analytically solvable Gaussian source models, with
and without space-momentum correlations, we present a comprehensive
overview of multi-boson symmetrization effects on particle
interferometry. For event ensembles of (approximately) fixed
multiplicity, the residual correlations lead to a minimum in the
correlation function at non-zero relative momentum, which can be
practically exploited to search, in a model-independent way, for 
multi-boson symmetrization effects in high-energy heavy-ion experiments. 
\end{abstract}

PACS number(s): 25.70-q, 25.70Pq, 25.70Gh

\section{Introduction}
\label{sec1}

In high energy hadronic and nuclear collisions large numbers of pions
are created (up to several thousand in Au+Au or Pb+Pb collisions at
the Relativistic Heavy Ion Collider (RHIC) or the Large Hadron
Collider (LHC)). This raises the issue of multi-particle
symmetrization effects among the final state particles and their
effects on the particle spectra and momentum-space correlations. If
the particles are set free from a state of sufficiently high
phase-space density, such effects are expected to become measurable.
This has triggered many investigations of multi-boson symmetrization
effects over the years (see
\cite{Zajc87,Pratt93,CGZ95,BK95,FW97,Zh97,ZC98,W98,RH98,ZSH98,AS99,BZ99,Letal00}
and further references given in those papers). In particular the 
possibility of Bose-Einstein condensation of pions in heavy-ion
collisions (sometimes misleadingly called ``pion laser'') has
attracted attention
\cite{Pratt93,CGZ95,ZC98,ZSH98,AS99,BZ99,Letal00}. Recent analyses of
experimental data point, however, to universally low pion
phase-space densities at freeze-out \cite{B94,E877_97,Fetal}. On the
other hand, the authors of Ref.~\cite{Rafelski} present a chemical
analysis of the hadronic final state in 158 $A$ GeV/$c$ Pb+Pb
collisions at the CERN Super Proton Synchrotron (SPS) which yields a
large pion chemical potential, $\mu_\pi\approx m_\pi$, corresponding
to rather large pion phase-space densities. The theoretical 
investigation of multi-boson symmetrization effects thus continues to
be a challenging problem of practical relevance.

In this paper we present a comprehensive ana\-ly\-sis of multi-boson 
symmetrization effects on multipar\-ticle spectra and Bose-Einstein 
interferometry. Our work not only provides a synthesis of many 
previously published investigations in a common and very
general formal framework, but also extends them in one important
aspect: while nearly all previous studies (the only exception 
known to us is Ref.~\cite{Zh99}) analyzed either event ensembles of
fixed particle multiplicity or an ensemble, in which multi-boson
symmetrization effects at finite phase-space density modified in a 
very specific way (see Sec.~\ref{sec3c}) a Poissonian input 
multiplicity distribution at vanishing phase-space density \cite{fn0}, 
our formalism allows to take into account arbitrary measured 
multipli\-ci\-ty distributions. 
In addition to the now well-established ``Bose clustering effect'', 
by which the effective emission function becomes narrower in 
{\em both coordinate and momentum space}, we show that multiboson 
symmetrization in general also introduces {\em ``residual correlations''} 
in the two-particle correlation function, i.e. deviations from unity 
which, in contrast to the Hanbury Brown--Twiss (HBT) effect resulting 
from two-particle exchange, are not directly related to the source 
size. These residual correlations modify the effective correlation 
strength $\lambda$, i.e. the ratio of the two-particle correlation 
function at zero and infinite relative momentum, and they introduce 
a new dependence of the correlator on the relative momentum $q$ 
which modifies the extraction of the source size via HBT radii.

These residual correlations are generic. We know only two multiplicity 
distributions for which they do not arise, the ``Poisson limit''
discussed in Sec.~\ref{sec3c} \cite{BK95,Zh97,ZC98,ZSH98,Letal00} 
and the ``covariant current formalism'' discussed in Appendix~\ref{app0} 
\cite{GKW,CH94}. In both cases, one prescribes an ``input''
multiplicity distribution (which refers to the infinite phase-space 
volume limit where multi-boson symmetrization effects can be neglected
\cite{fn0}), whereas the actually measured multiplicity distribution 
is a complicated function of the average phase-space density of the 
source. On the other hand, residual correlations do arise, in 
particular, for events with fixed multiplicity; there they lead to 
strong effects in sources with large phase-space density which were 
noted before \cite{Zajc87,Zh97,W98} but whose origin has up to
now not been fully clarified.  

The residual correlations and Bose clustering effects disappear for
dilute sources which emit particles from a state of low phase-space
density. Only in this limit the usual formalism (see, for example, the
recent reviews \cite{reviews}) for extracting effective source sizes
from two-particle correlation measurements can be applied. As the
source phase-space density (more accurately: the spatially averaged
phase-space density $\langle f\rangle(\bK)$
\cite{Letal00,B94,E877_97,Fetal} of particles with pair momentum  
$\bK$ at which the correlation function is evaluated) increases beyond 
a certain threshold, multi-boson effects set in rather abruptly. When
this happens, a more general formalism must be applied to separate
residual from HBT correlations. This is worked out in the paper. 

An important stepping stone for the discussion of multi-boson
symmetrization effects was provided 2 years ago by Cs\"org\H{o} and
Zim\'anyi \cite{ZC98} who found an analytic solution for the
corresponding recursion relations within a class of Gaussian source
models. This analytic solution dramatically reduces the complexity of
the problem and has thus triggered many recent (re-)investigations of
multi-boson effects; our paper is no exception. Before using them we 
rewrite their results \cite{ZC98} in a physically more instructive way 
which clearly exhibits the relevant physical parameters in the problem. 
This is straightforward, but we have not seen it published before, and 
it has important practical consequences. In the applications we focus
on Bose-Einstein interferometry studies of the two-particle correlation 
function; a detailed analysis of multi-boson effects on the measured
single-particle spectra was recently published in \cite{Letal00}. We
argue that, once the effects of multi-boson symmetrization on the 
normalization and shape of the two-particle correlation function are
qualitatively understood, they can be efficiently used to search
experimentally for multi-boson effects and for evidence of large
phase-space densities in high energy heavy ion collisions. This may
help to settle the controversy over whether or not pions are emitted
with large chemical potentials \cite{Fetal,Rafelski}.

Our paper is organized as follows: In Section \ref{sec2} we write down
a very general form of the density matrix for the ensemble of pion
emitting sources in terms of a superposition of elementary source
currents. The latter are assumed to have random phases to ensure
independent or ``chaotic'' particle emission. We introduce the source
Wigner density and the ``ring integrals'' \cite{Pratt93} from which
the normalization of the density operator and later the particle
spectra can be calculated. Carefully normalized expressions for the
latter are presented in Section \ref{sec3}, both for events with fixed
multiplicity and for multiplicity-averaged events. We define the
correlation strength and the incoherence parameter (the latter equals
1 for the fully chaotic sources studied in this paper), and we discuss
specifically the previously studied ``Poisson limit'' in which for
finite phase-space volumes multi-boson effects modify in a specific
way a Poissonian input multiplicity distribution at infinite
phase-space volume. In Section~\ref{sec4} we review and reformulate
the analytical solution \cite{ZC98} of the multi-boson recurrence
relations for Gaussian sources and show in particular that it can also
be applied to Gaussian sources with $x$-$p$-correlations (the Zajc
model \cite{Zajc93}) for which the momenta of the emitted particles
depend on the emission point in space-time. 

At this point the basic formalism is complete, and in Section
\ref{sec5} we proceed to discuss numerical results. We separate 
the two-particle correlation function into Bose-Einstein and 
residual correlations and study the asymptotic normalization of 
the correlator, the correlation strength, the HBT radii and the 
range of the residual correlations as a function of the pion 
multiplicity distribution, the pion phase-space density, and the
pair momentum, covering essentially the complete physical parameter
space. The numerical calculations are checked against analytic results
in the limits of large ($v\to\infty$) and small phase-space volume
($v\to 1$). For $v=1$ the source saturates the quantum mechanical
uncertainty limit, i.e. all pions occupy the same quantum state,
forming a Bose condensate. This interesting limit is under complete
analytical control. Experimental consequences are discussed in Section
\ref{sec6}: we argue that multi-boson effects can be detected by
triggering on fixed multiplicity and looking for a minimum at non-zero
relative momentum $q$ in the correlation function. Some concluding
remarks in Section \ref{sec7} close the paper.  

In three Appendices we provide technical details: Appendix \ref{app0}
discusses how our formalism is related to the ``covariant current
formalism'' of Gyulassy {\it et al.} \cite{GKW,CH94}, the only other
known ensemble which does not suffer from residual correlations in the
presence of multiboson effects. Appendices \ref{appa} and \ref{appb}
provide formulae for the analytic calculation of residual and
Bose-Einstein correlations in certain limits.  

\section{Source parametrization}
\label{sec2}

In this section we describe the parametrization of the density operator
on which our analysis of multi-boson effects will be based. Very little 
of the material in this section is original; the most important 
generalization of previous work, which almost exclusively discussed 
either fixed pion multiplicity or the case of a Poissonian multiplicity 
distribution in the absence of multi-boson symmetrization effects (the 
``Poisson limit'' of Sec.~\ref{sec3c}), is that we allow for an arbitrary 
pion multiplicity distribution. Still, although many of the relations 
given in this section can be found scattered in the literature, we find 
it useful to summarize them and thereby establish our notation and 
our normalization conventions. This last point is important since
much confusion over multiparticle symmetrization effects has arisen 
from insufficient attention to normalization issues. 

\subsection{The density operator}
\label{sec2a}

With the normalized (measured) pion multipli\-ci\-ty distribution 
$p_n$ we can write the density operator of the ensemble of collision 
events as
 \beq
 \label{1}
   \hat\rho = \sum_{n=0}^\infty p_n\, \hat\rho_n\,, \qquad
   \sum_{n=0}^\infty p_n =1 \, , \qquad
   \tr\,\hat\rho_n = 1\,,
 \eeq
where $\hat\rho_n$ is a normalized density operator for a system of 
exactly $n$ pions. We generalize the formalism developed in 
\cite{CGZ95,ZC98,GKW,CH94,SH97} and consider the ensemble as a 
superposition of localized source amplitudes $j_0$ whose number $N$ 
is distributed according to a normalized probability distribution $P_N$
(in general not related to $p_n$ above). The sources have identical
structure, but they are centered at different points $x_i$ and move
with different 4-velocities $u_i = \gamma_i(1,\bv_i)$. Their 
average positions and velocities are distributed by a normalized 
classical phase-space distribution $\rho(x_i,\bv_i)$:
 \beq
 \label{2}
   \int d\zeta_i \, \rho(\zeta_i) \equiv 
   \int d^4x_i \, \gamma_i^4\, d^3v_i\,  \rho(x_i,\bv_i) = 1\,.
 \eeq
Here we introduced the shorthand notation $\zeta_i$ (with the Lorentz 
invariant measure $d\zeta_i$) for the phase-space coordinates 
$(x_i,\bv_i)$ of the sources. ($\gamma_i^4\,d^3v_i = d^3u_i/u_i^0$ is
the invariant measure for the boost velocities.) Furthermore the source 
amplitudes $j_0$ are assumed to have a randomly distributed phase 
$\phi_i$. We can thus write for the total source \cite{CH94,SH97}
 \beq
 \label{3}
   J(x) = \sum_{i=1}^N e^{i\phi_i}\, j_0(\Lambda_i\cdot(x-x_i))
 \eeq
where $x_i$ is the source center in its rest frame and $\Lambda_i
 = \Lambda(\bv_i)$ denotes the Lorentz-boost into the global 
frame with the velocity $\bv_i$ of the elementary source $i$.
The on-shell Fourier transform ($p^0=E_p=\sqrt{m^2+\bp^2}$)
of $J$ is given by
 \bea
 \label{4}
   \tilde J(\bp) &=& \int d^4x\, e^{ip\cdot x} J(x)
   = \sum_{i=1}^N  e^{i\phi_i}\, e^{ip\cdot x_i}\,
   \tilde j_0(\Lambda_i\cdot p),
 \\
 \label{5}
    \tilde j_0(k) &=& \int d^4x\,  e^{ik\cdot x} j_0(x)\,.
 \eea
As a boosted on-shell momentum the argument of $\tilde j_0$ in (\ref{4})
is itself on-shell. We denote the on-shell normalization of $\tilde j_0$ 
by $n_0$:
 \beq
 \label{6}
   \int {d^3p\over E_p}\, \left\vert \tilde j_0(\bp) \right\vert^2 
   = n_0\,.
 \eeq
Following \cite{CGZ95,ZC98,GKW,CH94,SH97} we make the ansatz
 \beq
 \label{7}
  \hat\rho_n = {1\over{\cal N}(n)}
  \sum_{N=1}^\infty P_N
  \left(\prod_{i=1}^{N}\int d\zeta_i\, \rho(\zeta_i) 
  \int {d\phi_i\over 2\pi} \right) |n\ra \la n|
 \eeq
with
 \beq
 \label{8}
   \vert n \ra \equiv \vert n[N;\lbrace \zeta_i,\phi_i\rbrace]\ra
   = {1\over\sqrt{n!}}
     \left[ i\int {d^3p\over E_p} \tilde J(\bp) \hat a^+_p
     \right]^n \vert 0\ra .
 \eeq
$\hat a^+_p$ is the creation operator for a boson with on-shell momentum 
$\bp$ and satisfies the (covariantly normalized) commutation relations
 \beq
 \label{9}
    [\hat a_p,\hat a^+_{p'}] = E_p\, \delta(\bp-\bp')\,.
 \eeq
For convenience we denote from now on the Lorentz-invariant momentum-space 
integration measure $d^3p/E_p$ by $d\tp$. 

The assumed factorization in (\ref{7}) of the $n$-particle distribution 
$\rho(\zeta_1,\dots,\zeta_n)$ into a product of single-particle
phase-space distributions $\rho(\zeta_i)$ for the source current 
centers is often said to represent ``independent pion emission'' 
\cite{BK95}. In fact, this is only true if the states 
$|n\ra$ are orthogonal; as pointed out in \cite{ZC98} for the case 
studied here (where $J(x)$ describes a superposition of wave-packets 
$j_0$ localized at phase-space points $\zeta_i$), the emission of a 
pion by one current $j_0$ still depends on the positions of the
other current amplitudes, and if they are closely spaced in phase-space,
Bose-Einstein symmetrization leads to an emission probability which is
enhanced by their phase-space overlap (``stimulated emission'' 
\cite{ZC98}). -- The ansatz (\ref{7}) also provides a natural starting
point for generalizations to partially phase-coherent emission, by
introducing appropriate weight functions into the integrals over the
phases $\phi_i$ of the source currents \cite{Scotto}.

\subsection{Normalization of the density operator}
\label{sec2b}

The states $\vert n\ra$ are not normalized. Their norm
 \beq
 \label{10}
   \la n\vert n\ra = \bar n^n\,,\quad
   \bar n = \int d\tp\, |\tilde J(\bp)|^2 \,,
 \eeq
depends on the number $N$ of elementary source amplitudes $j_0$ and 
their phase-space positions $\zeta_i$ and phases $\phi_i$ (see (\ref{4})). 
If normalized states were used, this complicated expression would 
appear in the denominator of (\ref{7}) inside the integrals, thereby 
in general prohibiting the analytic evaluation of expectation values
\cite{ZC98}. The only known counterexample which, in spite of 
constructing the density operator from {\em normalized} states, 
allows for analytical evaluation of the spectra is the coherent state
formalism (``covariant current ensemble'', Appendix \ref{app0})
exploited in \cite{GKW,CH94,CGZ94}. On the other hand, the 
prescription (\ref{7},\ref{8}) not only permits analytic evaluation 
of the spectra, but also correctly implements stimulated emission 
of bosons, ensuring that for thermalized systems the Bose-Einstein 
single-particle distribution is recovered in the infinite volume 
limit \cite{ZC98,Scotto,Slotta}. 

The normalization of $\hat\rho_n$ therefore must be ensured explicitly
by adding the factor $1/{\cal N}(n)$ in (\ref{7}). The latter can be 
calculated by using
 \beq
 \label{11}
   \int {d\phi_i\over 2\pi}   \int {d\phi_j\over 2\pi}\,
   e^{i\phi_i}\, e^{-i\phi_j} = \delta_{ij},
 \eeq
keeping only the leading terms in $N$ \cite{GKW,CH94}:
 \bea
 \label{12}
   &&1 = \tr\,\hat\rho_n \approx
   {1\over{\cal N}(n)}  \sum_{N=1}^\infty P_N\, N(N-1)\cdots(N-n+1)
 \nonumber\\
   &&\qquad\qquad\ \times
   \int d\tp_1\cdots d\tp_n\, S_n(\bp_1,\dots,\bp_n)\, ,
 \\
 \label{13}
   && S_n(\bp_1,\dots,\bp_n) = n_0^n\,\sum_{\sigma^{(n)}} 
   \bar\rho(\bp_1,\bp_{\sigma_1})\cdots \bar\rho(\bp_n,\bp_{\sigma_n}).
 \eea
Here $\sigma^{(n)}$ denotes the set of permutations of the indices 
$(1,\dots,n)$, and
 \bea
 \label{14}
  \bar\rho(\bp_i,\bp_j) &=& {1\over n_0}
  \int d^4x \, \gamma_i^4\, d^3v \, \rho(x,\bv)\, e^{-i(p_i-p_j)\cdot x}
 \nonumber\\
   &&\times \, \tilde{j_0}^*\Bigl(\Lambda(\bv)\cdot p_i\Bigr)\,
               \tilde{j_0}\Bigl(\Lambda(\bv)\cdot p_j\Bigr)
 \eea
is the normalized two-particle exchange amplitude:
 \beq
 \label{15}
   \int d\tp\, \bar\rho(\bp,\bp) = 1\,.
 \eeq
The approximation in (\ref{12}) neglects the $N^n-N(N-1)\cdots(N-n+1)$
terms in which more than one pion was emitted from the same source $j_0$;
their relative contribution is of order
 \beq
 \label{16}
   {N^n-N(N-1)\cdots(N-n+1)\over N^n} 
   \approx {n(n-1)\over 2 N}
 \eeq
and can be neglected if the mean number of sources is much larger than
the average number of emitted pairs: 
 \beq
 \label{17}
   \lda N\rda 
   \gg
   \half \la n(n-1)\ra 
   \,.
 \eeq   
(We denote by $\la\dots\ra$ averages with respect to the pion multiplicity
distribution $p_n$ and by $\lda\dots\rda$ averages with respect to $P_N$.)
In this limit we thus have
 \bea
 \label{18}
   {\cal N}(n) &=& \lda N(N-1)\cdots(N-n+1)\rda
 \nonumber\\
   &&\times
   \int d\tp_1\cdots d\tp_n\, S_n(\bp_1,\dots,\bp_n)\, .
 \eea
The limit (\ref{17}) can be enforced by letting in (\ref{6}) $n_0\ll 1$, 
i.e. each individual source $j_0$ emits on average much less than 1 pion,
and the total emission process is composed of a very large number of 
elementary emission processes. In Sec.~\ref{sec4} below we will see that 
the parameter $n_0$ drops out from the expressions for the spectra and 
correlation functions and thus is physically irrelevant. Its sole purpose 
in the formalism is to ensure the applicability of the approximations 
(\ref{17}) and (\ref{12}). The corresponding idealization of the emitter 
as a superposition of very many weak source currents should be a good 
approximation for high-energy heavy-ion collisions, where each event 
produces anyway a large number of hadrons. It may, however, require 
reconsideration for collisions between elementary particles (e.g. 
$e^+e^-$ collisions at LEP), where only a handful of identical pions 
is created per collision such that the additional smoothening implied 
by setting $n_0\ll 1$ may be less innocuous. 

\subsection{The ring algebra}
\label{sec2c}

${\cal N}(n)$ is related to the ``sum of ring integrals'' defined in 
\cite{Pratt93},
 \beq
 \label{19}
   \omega(n) = {1\over n!} 
   \int d\tp_1\cdots d\tp_n\, S_n(\bp_1,\dots,\bp_n)\, ,
 \eeq
by 
 \beq
 \label{20}
   {\cal N}(n) = n!\, \lda N(N-1)\cdots(N-n+1)\rda \, \omega(n)\,.
 \eeq
The evaluation of $\omega(n)$ is done recursively using Pratt's 
``ring algebra'' \cite{Pratt93}. One defines \cite{Pratt93,CGZ95,ZC98} 
 \bea
 \label{21}
    G_i(\bp,\bq) &=& n_0^i \int \bar\rho(\bp,\bk_1)\,d\tk_1\,
    \bar\rho(\bk_1,\bk_2)\cdots d\tk_{i-1} 
    \bar\rho(\bk_{i-1},\bq)
 \nonumber\\
    &=& \int G_1(\bp,\bk)\,d\tk \ G_{i-1}(\bk,\bq)
 \eea
with $G_1(\bp,\bq)=n_0\,\bar\rho(\bp,\bq)$ and normalization 
 \beq
 \label{22}
    C_i = \int d\tp \ G_i(\bp,\bp)
 \eeq
(with $C_1=n_0$). Identification in (\ref{21}) of $\bp$ with $\bq$ 
defines a ``ring''; for this reason (\ref{22}) is sometimes called a 
``ring integral''. For the sum of ring integrals $\omega(n)$ one
finds the recursion relation \cite{CGZ95}
 \beq
 \label{23}
   \omega(n) = {1\over n} \sum_{i=1}^n C_i\, \omega(n-i)\,,
 \eeq
with $\omega(1)=n_0$ and the definition $\omega(0)=1$.

\subsection{The source Wigner density}
\label{sec2d}

The two-particle exchange amplitude $\bar\rho(\bp_i,\bp_j)$ can be 
rewritten in terms of the source Wigner density $g(x,K)$:
 \bea
 \label{24}
   \bar\rho(\bp_i,\bp_j) &=&
   \int d^4x\, g\left(x,\half(p_i+p_j)\right)\,
   e^{-i(p_i-p_j)\cdot x}
 \nonumber\\
   &\equiv& \int d^4x\, g\left(x,K_{ij}\right)\,
   e^{-i q_{ij}\cdot x}\, .
 \eea
Here we introduced the average $K_{ij}$ and the difference $q_{ij}$ of 
the two on-shell momenta $p_i,p_j$. $g(x,K)$ is the quantum mechanical 
analogue of the phase-space distribution of the emitted pions. 
According to (\ref{14}) it is a folding of the classical phase-space 
distribution $\rho$ of the elementary source centers with their 
individual Wigner densities \cite{CH94}:
 \bea
 \label{25}
   g(x,K) &=& \int d^4z\,\gamma^4 d^3v\,\rho(z,\bv)\, 
   g_0\bigl(x{-}z,\Lambda(\bv)\cdot K\bigr) ,
 \\
 \label{26}
   g_0(x,p) &=& {1\over n_0} \int d^4y \, e^{-ip\cdot y}\, 
   j_0^*\left(x+{\textstyle{y\over2}}\right)\,
   j_0\left(x-{\textstyle{y\over2}}\right)\,.
 \eea
$g$ and $g_0$ are normalized to 1:
 \beq
 \label{27}
   \int d\tp\ d^4x\, g(x,p) = \int d\tp\ d^4x\, g_0(x,p) = 1\,.
 \eeq

\section{Multiparticle distributions}
\label{sec3}

The way in which multi-boson symmetrization effects manifest 
themselves in Bose-Einstein interferometry differs for ensembles 
of events with fixed multiplicity \cite{Zajc87,Zh97,W98} and for 
multiplicity-averaged ensembles \cite{Pratt93,CGZ95,Zh97,ZC98,ZSH98}. 
In this section we provide the relevant formulae for both cases.
In the multiplicity-averaged case (Sec.~\ref{sec3b}) we allow for 
arbitrary pion multiplicity distributions; the previously studied 
specific case of Poisson statistics in the dilute gas limit 
\cite{BK95,Zh97,ZC98,ZSH98,Letal00} is shortly reviewed in
Sec.~\ref{sec3c}. In Sec.~\ref{sec3d} we define and differentiate
between the correlation strength and the incoherence parameter and
point out that all sources studied here are fully chaotic even if the
correlation strength differs from 1.

\subsection{Events with fixed pion multiplicity}
\label{sec3a}

We first discuss the case of fixed pion multiplicity $n$,
$p_m=\delta_{mn}$. With the ingredients from Sec.~\ref{sec2} 
the Lorentz-invariant $n$-pion distribution in an ensemble of
events containing exactly $n$ pions each can be written as
 \bea
 \label{28}
   N_n^{(n)}(\bp_1,\dots,\bp_n) &=&
   E_1\cdots E_n \tr\left( \hat\rho_n\, 
   \hat a^+_{p_1}\cdots \hat a^+_{p_n} \hat a_{p_n}\cdots \hat a_{p_1}
   \right)
 \nonumber\\
   &=& {n!\, S_n(\bp_1,\dots,\bp_n) \over
        \int d\tp_1\cdots d\tp_n\, S_n(\bp_1,\dots,\bp_n)}\,.
 \eea
It is normalized to $n!$. The $i$-particle distribution in an $n$-pion 
state is (for $i\leq n$ and in the limit (\ref{17})) given 
by
 \bea
 \label{29}
   &&N_i^{(n)}(\bp_1,\dots,\bp_i) =
   E_1\cdots E_i\, \tr\left( \hat\rho_n\, 
   \hat a^+_{p_1}\cdots \hat a^+_{p_i} \hat a_{p_i}\cdots \hat a_{p_1}
   \right)
 \nonumber\\
   &&\qquad = {n!\over(n-i)!} \int d\tp_{i+1}\cdots d\tp_n\,
        N_n^{(n)}(\bp_1,\dots,\bp_n)
 \nonumber\\
   &&\qquad = {n!\over(n-i)!} 
       {\int d\tp_{i+1}\cdots d\tp_n\,S_n(\bp_1,\dots,\bp_n) 
        \over
        \int d\tp_1\cdots d\tp_n\, S_n(\bp_1,\dots,\bp_n)}\,.
 \eea
It is normalized to $n(n{-}1)\cdots(n{-}i{+}1)$. Note that in the
limit (\ref{17}) the dependence on the number of sources, $N$, and 
its distribution, $P_N$, drops out completely. This differs from the 
result obtained in \cite{CH94} using the so-called covariant current
ensemble \cite{GKW}. The latter is based on a density operator which
is constructed from a superposition of normalized coherent states. 
In the covariant current ensemble the one- and two-particle inclusive
spectra (averaged over the pion multiplicity distribution predicted by
that model) were found to depend explicitly on $N$ and $P_N$. This
is further discussed in Appendix \ref{app0}.

For the one- and two-particle distributions in an $n$-pion event 
Eq.~(\ref{29}) reduces to \cite{Pratt93,CGZ95}
 \bea
 \label{30}
   N^{(n)}_1(\bp_1) &=& n \cdot 
   {\int d\tp_2\cdots d\tp_n\, S_n(\bp_1,\dots,\bp_n)
    \over 
    \int d\tp_1\cdots d\tp_n\, S_n(\bp_1,\dots,\bp_n)}
 \nonumber\\
   &=& \sum_{i=1}^{n} {\omega(n-i)\over \omega(n)} G_i(\bp_1,\bp_1)\, ,
 \eea
and 
 \bea
 \label{31}
   &&N_2^{(n)}(\bp_1,\bp_2) = n(n-1) \cdot
   {\int d\tp_3\cdots d\tp_n\, S_n(\bp_1,\dots,\bp_n)
    \over 
    \int d\tp_1\cdots d\tp_n\, S_n(\bp_1,\dots,\bp_n)}
 \nonumber\\
   && = \sum_{i=2}^{n} {\omega(n-i)\over \omega(n)} 
        \sum_{j=1}^{i-1} \Bigl[G_j(\bp_1,\bp_1)\,G_{i-j}(\bp_2,\bp_2)
 \nonumber\\
   && \qquad\qquad\qquad\qquad
   +\, G_j(\bp_1,\bp_2)\,G_{i-j}(\bp_2,\bp_1)\Bigr].
 \eea

The 2-particle correlation function from $n$-pion events is given 
by the ratio
 \beq
 \label{32}
   C_2^{(n)}(\bp_1,\bp_2) = C_2^{(n)}(\bq,\bK) = 
   {N_2^{(n)}(\bp_1,\bp_2) \over  N^{(n)}_1(\bp_1)\, N^{(n)}_1(\bp_2)} \,,
 \eeq
where $\bq=\bp_1-\bp_2$ and $\bK=(\bp_1+\bp_2)/2$. 

\subsection{Multiplicity-averaged $n$-particle spectra}
\label{sec3b}

Multiplicity-averaged expressions are now easily derived by
averaging with the pion multiplicity distribution $p_n$ according 
to Eq.~(\ref{1}). For the $n$-pion inclusive spectrum one finds
 \bea
 \label{34}
   N_i(\bp_1,\dots,\bp_i) &=& 
   E_1\cdots E_i\, {dN\over d^3p_1\cdots d^3p_i}
 \nonumber\\   
   &=&  E_1\cdots E_i\, \tr\left( \hat\rho\, 
   \hat a^+_{p_1}\cdots \hat a^+_{p_i} \hat a_{p_i}\cdots \hat a_{p_1}
   \right)
 \nonumber\\
   &=& \sum_{n=i}^\infty p_n\, N_i^{(n)}(\bp_1,\dots,\bp_i)\, .
 \eea 
It is normalized to $\la n(n-1)\cdots(n-i+1)\ra$ (i.e. the $i^{\rm th}$ 
factorial moment). Inserting the expressions (\ref{30}) and (\ref{31}), 
interchanging the summations over $n$ and $i$, defining
 \beq
 \label{35}
   h_i = \sum_{n=i}^\infty p_n\, {\omega(n-i)\over\omega(n)}\, ,
 \eeq
and using the methods of Refs.~\cite{Pratt93,CGZ95} one finds the 
following simple expressions:
 \bea
 \label{36}
 \FL   
   &&E_p {dN\over d^3p} = N_1(\bp) = \sum_{i=1}^\infty h_i\, 
     G_i(\bp,\bp)\, ,
 \\
 \label{37}
   &&E_1 E_2 {dN\over d^3p_1\, d^3p_2} = N_2(\bp_1,\bp_2)
 \nonumber\\
   && \ \
     = \sum_{i,j=1}^\infty h_{i+j}\, \Bigl[
        G_i(\bp_1,\bp_1)\, G_j(\bp_2,\bp_2)
 \\
   &&\qquad\qquad\qquad
   + G_i(\bp_1,\bp_2)\, G_j(\bp_2,\bp_1)\Bigr]\, .
 \nonumber
 \eea
For the two-particle correlation function one obtains
 \bea
 \label{38}
   &&C_2(\bp_1,\bp_2)=
 \\
   && {\sum_{i,j} h_{i+j} 
       \bigl[G_i(\bp_1,\bp_1) G_j(\bp_2,\bp_2){+}G_i(\bp_1,\bp_2)
                              G_j(\bp_2,\bp_1)\bigr]
       \over
       \sum_{i,j} h_i\, h_j\, G_i(\bp_1,\bp_1)
                                       G_j(\bp_2,\bp_2)}\, .
 \nonumber
 \eea
The corresponding expressions in the previous subsection are
recovered by setting $p_m=\delta_{mn}$ (i.e. by substituting 
$h_i=0$ for $i>n$ and $h_i = {\omega(n-i)\over\omega(n)}$ for 
$i\leq n$).

\subsection{``Poisson limit''}
\label{sec3c}

These expressions simplify considerably \cite{CGZ95,Zh97} if the 
following special multiplicity distribution is assumed:
 \beq
 \label{40}
   \bar p_n = {\omega(n)\,\bar N^n
         \over \sum_{k=0}^\infty \omega(k)\,\bar N^k}\, .
 \eeq
Here $\bar N$ is an arbitrary scaling parameter which leaves the 
structure of the theory unchanged. The distribution $\bar p_n$ 
depends on $v$ since the sum of ring integrals $\omega(n)$ depends 
on $v$. In the limit of a large phase-space volume $v$ (i.e. for a 
very dilute and sufficiently hot pion gas) the higher order ring 
integrals $C(i)$, $i>1$, can be neglected, and $\omega(n)$ 
approaches \cite{Pratt93,CGZ95,Zh97}
 \beq
 \label{41}
   \lim_{v\to\infty} \omega(n) = {n_0^n\over n!}\,.
 \eeq
In this limit $\bar p_n$ becomes a Poisson distribution with mean 
multiplicity $\la n \ra = n_0\bar N$:
 \beq
 \label{42}
   \lim_{v\to\infty} \bar p_n =
   \lim_{v\to\infty} {\omega(n)\,\bar N^n
   \over \sum_{k=0}^\infty \omega(k)\,\bar N^k}
   = {\la n\ra \over n!}\, e^{-\la n \ra}\,.
 \eeq
This is why we call this special case the ``Poisson limit'', with
quotation marks to indicate that the multiplicity distribution
is Poissonian only in the limit $v\to\infty$. In the opposite 
limit $\bar p_n$ becomes a Bose-Einstein distribution with mean 
multiplicity $\la n\ra = n_0\bar N/(1{-}n_0\bar N)$ \cite{ZC98}:
 \beq 
 \label{42a}
   \lim_{v\to 1} \bar p_n = 
   {\la  n \ra^n \over \left(1+\la  n \ra\right)^{n+1}}\,.
 \eeq
This can be easily confirmed by using the limiting expressions derived 
in the next section.

For the choice (\ref{40}) the factors $h_i$ in (\ref{35}) are given by
$h_i=\bar N^i$ and satisfy $h_{i{+}j}=h_i h_j$. Eqs.~(\ref{36})--(\ref{38}) 
then reduce to
 \bea
 \label{43}
   N_1(\bp) &=& H(\bp,\bp)\, ,
 \\
 \label{44}
   N_2(\bp_1,\bp_2) &=& H(\bp_1,\bp_1) H(\bp_2,\bp_2)
 \nonumber\\
                    && + H(\bp_1,\bp_2) H(\bp_2,\bp_1)\, ,
 \\
 \label{45}
   C_2(\bp_1,\bp_2) &=& 1 
   + {H(\bp_1,\bp_2) H(\bp_2,\bp_1) \over
      H(\bp_1,\bp_1) H(\bp_2,\bp_2)}\, ,
 \eea
with the auxiliary function 
 \beq
 \label{51}
   H(\bp,\bq) = \sum_{i=1}^{\infty} \bar N^i\,G_i(\bp,\bq)\,.
 \eeq
With the ``effective emission function'' $S(x,\bK)$ defined by \cite{ZSH98}
 \beq
 \label{52}
   H(\bp_1,\bp_2) = \int d^4x\, S(x,\bK)\, e^{iq\cdot x}
 \eeq
the correlation function can then be written in the ``standard form'' 
\cite{CH94,reviews,S73}
 \bea
 \label{53}
   C_2(\bq,\bK) &=& 1 + {\left\vert\int d^4x\, S(x,\bK)\, 
                     e^{iq\cdot x}\right\vert^2 \over
   \int d^4x\, S(x,\bp_1) \int d^4y\, S(x,\bp_2)}
 \\
       &=& 1 + {[N_1(\bK)]^2\over
                     N_1(\bp_1) N_1(\bp_2)}\, 
   \left\vert {\int d^4x\,S(x,\bK)\, e^{iq{\cdot}x}\over
               \int d^4x\, S(x,\bK)}\right\vert^2 .
 \nonumber
 \eea
Note that this simplification occurs only for this particular
multiplicity distribution (which in this paper we will call 
``the Poisson limit'' for shortness). It allows relatively easy
access to the space-time structure of the effective emission 
function, via the last factor in the second term of (\ref{53}) 
which can be isolated from the measured correlation function
by subtracting the 1 and dividing by the (measured) ratio of 
single particle spectra $[N_1(\bK)]^2/[N_1(\bp_1) N_1(\bp_2)]$. 
For other multiplicity distributions we will see that the 
space-time interpretation of the correlation function is less 
direct and more complicated.

As far as we are aware, except for \cite{Zh99} all previous studies 
of multiparticle symmetrization effects refer either to the case of 
fixed multiplicity or to the above ``Poisson limit''.

\subsection{Correlation strength and incoherence parameter}
\label{sec3d}

Since the quantum statistical correlations introduced by multiparticle 
symmetrization effects are expected to disappear for large relative 
momenta, it seems reasonable to define the {\em correlation strength} 
$\lambda(\bK)$ by \cite{Zajc87}
 \beq
 \label{33}
   \lambda(\bK) = \lim_{\vert\bq\vert\to\infty}
   {C_2(\bbox{0},\bK)\over  C_2(\bq,\bK)} - 1\,.
 \eeq
In the past this correlation strength has often been identified with 
the degree of chaoticity or lack of phase coherence of the particle 
source: sources which emit particles coherently were taken to yield 
$\lambda=0$ while fully chaotic sources correspond to $\lambda=1$. As 
we will see below, this is true only in the ``Poisson limit'' of the 
preceding subsection, but not for general multiplicity distributions 
where the correlation strength (\ref{33}) can become either much larger 
than 1 or even negative. What remains true for arbitrary multiplicity 
distributions is that for a coherent source the se\-cond (2-particle 
exchange) terms in the square brackets of Eqs.~(\ref{31}) and (\ref{37})
vanish, and that for partially cohe\-rent sources (like those discussed in 
\cite{APW,Biya,Cramer,HZ97}) they amount to only a fraction of the first 
term even at $\bq=0$. Partially coherent sources can thus be characterized 
by an {\em incoherence parameter} $0 \leq \lambda_{\rm incoh}(\bK) < 1$ 
where $\lambda_{\rm incoh}(\bK)$ is defined by
 \beq
 \label{incoh}
   \lambda_{\rm incoh}(\bbox{K}) = {N_2^{\rm sym}(\bK,\bK)
   \over N_2^{\rm unsym}(\bK,\bK)} - 1\, ,
 \eeq
with $N_2^{\rm sym}$ denoting the full, symmetrized expressions 
(\ref{31}), (\ref{37}), and $N_2^{\rm unsym}$ denoting only the first 
terms inside the square brackets, without the exchange term, 
respectively. 

Since the first and second terms in (\ref{31}) and (\ref{37}) are always 
equal to each other at $\bq=0$, the sources discussed in the present 
paper correspond to 
 \beq
 \label{incoh2}
   \lambda_{\rm incoh}(\bK) = 1\, ,
 \eeq
i.e. to {\em fully chaotic sources}, for all $\bK$ and {\em independent 
of the multiplicity distribution}. This statement applies also to the 
``Poisson limit'' discussed in the preceding subsection, in particular 
to the limit in which the parameter $n_0$ approaches the critical value 
$n_c$ for Bose-Einstein condensation as discussed in
\cite{ZC98,ZSH98,Letal00}. Similar to the case of an ideal Bose gas 
in statistical mechanics, {\em Bose-Einstein condensation is thus not
  correlated with the onset of phase coherence in the source.} As also 
pointed out in \cite{Letal00}, this differs from the situation in a
laser in which the multi-boson state is a single coherent state which 
exhibits complete phase coherence, shows no Bose-Einstein
correlations, and features a Poissonian multiplicity distribution. 

Eq.~(\ref{53}) shows that in the ``Poisson limit'' also the correlation 
strength (\ref{33}) is $\lambda(\bK) = 1$ independent of $\bK$, and 
that in this limit it indeed agrees with the incoherence parameter,
$\lambda_{\rm incoh}$.
 
\section{Gaussian sources}
\label{sec4}

Due to their analytic tractability Gaussian source mo\-dels have
always been popular in Bose-Einstein interfero\-me\-try (for recent 
reviews see \cite{reviews}). It was recently discovered \cite{ZC98}
that they also allow for an analytic evaluation of multi-particle 
symmetrization effects. Since then se\-ve\-ral investigations exploiting 
this analytic solution have appeared \cite{Zh97,ZSH98,Zh99} which 
provide a useful basis for comparison. For this reason we adopt 
the same model here.

The results of Secs.~\ref{sec2} and \ref{sec3} show that the 
multiparticle spectra are fully determined once the pion multipli\-ci\-ty
distribution $p_n$ has been specified and the functions 
$G_i(\bp_1,\dots,\bp_i)$ have been calculated for all $i$.
From the definition of the latter, Eq.~(\ref{21}), we see that
the $G_i$ are completely controlled by the structure of the 
two-particle exchange amplitude $\bar\rho(\bp,\bq)$ or, equivalently,
of the source Wigner density $g(x,\bK)$ (see (\ref{24})).
That the latter can (according to (\ref{25})) be thought of as 
composed of {\em two} ingredients, the elementary source currents 
$j_0$ and the phase-space distribution $\rho(\zeta)$ of their 
centers, does not matter in practice, i.e. it has no observable 
con\-sequences \cite{CRIS98,WFH99}. For a given source Wigner density 
$g(x,\bK)$, equation (\ref{25}) in general allows for infinitely many 
different decompositions into a classical phase-space distribution 
$\rho$ and an elementary source Wigner density $g_0$. This becomes
manifest in the case of nonrelativistic Gaussians (studied e.g. in 
\cite{ZC98,W98,CRIS98,WFH99,MP,Wetal,Geiger}) where the Gaussian widths 
of $g(x,\bK)$ in coordinate and momentum space, $R$ and $\Delta$,
are related to those characterizing $\rho$ ($R_0$ and $\Delta_0$) and
$g_0$ ($\sigma/\sqrt{2}$ and $1/(\sqrt{2}\sigma)$ --- in those papers 
$g_0$ was constructed from minimum uncertainty wave packets of size 
$\sigma$) by
 \beq
 \label{54}
    R^2=R_0^2+{\sigma^2\over 2}\, , \qquad
    \Delta^2 = \Delta_0^2 + {1\over 2\sigma^2}\, .
 \eeq
Obviously this has, for given $R$ and $\Delta$, an infinity of solutions
$(R_0,\Delta_0,\sigma)$. Therefore, the wave packet width $\sigma$
is not directly measurable and can at most be bounded by the two
measurable quantities $R$ and $\Delta$.

\subsection{Static sources}
\label{sec4a}

We will here study the non-relativistic Gaussian source Wigner density
 \beq
 \label{55}
   g(x,\bp) = {E_p\, \delta(x^0)\over (2\pi R\Delta)^3}\, 
   \exp\left( - {\bx^2\over 2R^2}
              - {\bp^2\over 2\Delta^2}\right)\, ,
 \eeq
which describes instantaneous emission of particles with a Gaussian
(non-relativistic thermal) momentum distribution from a sphere with a
Gaussian density distribution. The momenta do not depend on position,
i.e. $\bx$ and $\bp$ are uncorrelated. For the source (\ref{55}) the 
two-particle exchange amplitude reads 
 \beq
 \label{56}
   \bar\rho(\bp_1,\bp_2) = 
   {\sqrt{E_1\,E_2}\over (2\pi\Delta^2)^{3\over 2}}\,
   \exp\left( - {\bK^2\over 2\Delta^2} 
              - {R^2\,\bq^2\over 2}\right)\,.
 \eeq
The higher-order exchange amplitudes $G_n(\bp_1,\bp_2)$, $n\geq 1$, 
can be calculated analytically \cite{ZC98} from recursion relations 
\cite{CGZ95} gene\-ra\-ted by Eq.~(\ref{21}):
 \begin{mathletters}
 \label{57}
 \bea
 \label{57a}
 \FL
   &&{G_n(\bp_1,\bp_2)\over \sqrt{E_1\,E_2}} =
   {n_0^n\, c_n\over (2\pi\Delta^2)^{3\over 2}}\,
   \exp\left( - {\bK^2\over 2\Delta_n^2} 
              - {R_n^2\,\bq^2\over 2}\right),
 \\
 \label{57b}
   &&R_n^2 = a_n\,R^2\, ,\qquad \Delta_n^2 = a_n\,\Delta^2\, ,
 \\
 \label{57c}
   &&a_n = {1\over v}\, {(v+1)^n + (v-1)^n \over (v+1)^n - (v-1)^n}
   \leq 1\, , 
 \\
 \label{57d}
   &&c_n = \left( 2^{2n}\, v \over (v+1)^{2n} -(v-1)^{2n} 
                \right)^{3/2} \leq 1\, ,
 \eea
 \end{mathletters}
where 
 \beq
 \label{58}
   v = 2R\Delta \geq 1
 \eeq
is a measure for the total phase-space volume occupied by the
source Wigner function $g(x,\bK)$. (The latter would be given by 
$v^3$, and $v{=}1$ corresponds to the minimum value allowed by the 
uncertainty relation.) The ring integrals $C_n$ are given by
 \beq
 \label{59}
   C_n = c_n\, a_n^{3\over 2}\, n_0^n\, .
 \eeq
It is easy to generalize expressions (\ref{55})--(\ref{58}) to the 
case of a Gaussian source with different widths $R_1,R_2,R_3$ and 
$\Delta_1,\Delta_2,\Delta_3$ in the three spatial directions, with 
corresponding phace-space volume factors $v_1,v_2,v_3$.

From the above results one sees that the parameter $n_0$ (the 
normalization (\ref{6}) of the elementary sources) drops out from 
all observables: Eqs.~(\ref{59}), (\ref{23}) and (\ref{35}) imply 
that $\omega(n)\sim n_0^n$ and $h_i\sim n_0^{-i}$. Since also $G_i\sim
n_0^i$, all factors of $n_0$ cancel in Eqs.~(\ref{30}), (\ref{31}),
and (\ref{36})--(\ref{38}). {\em We will therefore, without loss of 
generality, set $n_0=1$ from now on} \cite{fn}.

In Eqs.~(\ref{57}) we have rewritten the results of Zim\'anyi and 
Cs\"org\H o \cite{ZC98} in such a way that the essential structure 
of the multiparticle symmetrization effects becomes ma\-ni\-fest:
first, Eq.~(\ref{57a}) shows that $n$-body symmetrization effects 
contribute to the multiparticle spectra with relative weight $c_n{\leq}1$, 
and these contributions correspond to effective source functions 
with modified widths $R_n{\leq}R$ and $\Delta_n{\leq}\Delta$. This 
reflects the fact that bosons like to cluster in phase-space. Second, 
the weights $c_n$ and the strength of the clustering, given by $a_n$, 
are controlled {\em by a single parameter}, the available phase-space 
volume $v$. 

For large $v$ multiparticle effects are weak; one finds for $n\geq 1$
 \begin{mathletters}
 \label{60}
 \bea
 \label{60a}
   \lim_{v\to\infty} a_n &=& {1\over n}\, ,
 \\
 \label{60b}
   \lim_{v\to\infty} c_n &=& {1\over n^{3\over 2}}
   \left( {2\over v} \right)^{3(n-1)}\, ,
 \\
 \label{60c}
   \lim_{v\to\infty} C_n &=& {1\over n^3} 
   \left( {2\over v}\right)^{3(n-1)}\,,
 \\
 \label{60d}
   \lim_{v\to\infty} \omega(n) &=& {1\over n!}\,.
 \eea
 \end{mathletters}
In this limit the weight of $n^{\rm th}$-order symmetrization effects
is suppressed by the $n-1^{\rm st}$ power of the {\em input} phase-space 
density $d_0=1/v^3$ (i.e. the phase-space density in the absence of 
multi-boson effects). 

As $v$ becomes small and approaches the uncertainty limit, 
$v{\to}1{+}\epsilon$, one finds for $n\geq 2$ 
 \begin{mathletters}
 \label{61}
 \bea
 \label{61a}
   \lim_{v\to 1+\epsilon} a_n &=& 1-\epsilon  ,
   \ \ 
   \lim_{v\to 1+\epsilon} {a_{n+1}\over a_n} = 
   1-2\left({\epsilon\over2}\right)^n  ,
 \\
 \label{61b}
   \lim_{v\to 1+\epsilon} c_n &=& 1-{\textstyle{3\over 2}}(n{-}1)\epsilon\,,
 \\
 \label{61c}
   \lim_{v\to 1+\epsilon} C_n &=& 1-{\textstyle{3\over 2}}n\epsilon\,,
 \\
 \label{61d}
   \lim_{v\to 1+\epsilon} \omega(n) &=& {\lim_{v\to 1+\epsilon}}c_n
   = 1-{\textstyle{3\over 2}}(n{-}1)\epsilon\,.
 \eea
 \end{mathletters}
(Of course, $a_1{=}c_1{=}C_1{=}1$ independent of $v$.) Now all 
$n$-boson terms contribute with equal weight. However, the bosonic 
clustering effects become weak, $a_n{\to}1$, since the source is 
already as small as allowed by the uncertainty principle; all 
$n$-boson terms thus reflect the same effective source parameters. 
In the following section these features and their implications for 
Bose-Einstein interferometry will be studied quantitatively.

\subsection{Gaussian sources with $x$-$p$-correlations}
\label{sec4b}

In this subsection we shortly describe a Gaussian model, first
suggested by Zajc \cite{Zajc93}, in which the momenta of the emitted
particles are correlated with the emission points, as is the case
e.g. in hydrodynamically expanding sources. The model allows to
control the strength of the $x$-$p$-correlations by tuning a parameter
$s$ and thus presents a valuable playground to investigate the
qualitative effects of $x$-$p$-correlations on HBT interferometry. We
will show that, for the purposes of studying multi-boson effects, it
can be completely mapped onto the static Gaussian source presented in
the previous subsection: after a suitable redefinition of the physical 
parameters, the algebra for calculating multi-boson effects becomes
identical to the already solved static problem.
 
Zajc's model starts from the non-relativistic classical phase-space
distribution
 \bea
 \label{n1}
   \rho(x,\bp) &=& {\delta(x^0)\over (2\pi R_s\Delta_0)^3}
 \\
   &&\times
   \exp\left[ -{1\over 2(1{-}s^2)}\left(
   {\bx^2\over R_0^2} - 2s{\bx\cdot\bp\over R_0\Delta_0}
   + {\bp^2\over\Delta_0^2}\right)\right]\,,
 \nonumber
 \eea
where
 \beq
 \label{n2}
   R_s = R_0\sqrt{1{-}s^2}, \ \  0\leq s\leq 1\, \ \ 
   {\rm and} \ \ \bp = m\bv.  
 \eeq
The term proportional to $s$ in the exponent couples momentum to
position; in the limit $s\to 1$ one obtains perfect 
$x$-$p$-correlations:
 \beq
 \label{n3}
   \lim_{s\to 1} \rho(x,\bp) \sim \delta(x^0)\, 
   \delta\left({\bx\over R_0}-{\bp\over \Delta_0}\right)\,.
 \eeq
This violates, of course, the uncertainty relation; the threshold at 
which the effective phase-space volume factor $v_s=2R_s\Delta_0$
becomes smaller than 1 is given by \cite{Geiger}
 \beq
 \label{n4}
   s^2_{\rm crit} = 1 - (2R_0\Delta_0)^{-2}\,.
 \eeq

To obtain a valid source Wigner density which does not violate the 
uncertainty relation, we now fold \cite{CH94} $\rho$ according to 
(\ref{25}) (leaving out the factor $\gamma^4$ in the integration 
measure since the source is non-relativistic) with the Wigner density 
of the elementary source currents $j_0$, which are taken as 
non-relativistic minimum uncertainty Gaussian wave-packets of width 
$\sigma$:
 \beq
 \label{n5}
   g_0(x,\bp) = {E_p\,\delta(x^0)\over \pi^3}\,
   \exp\left[-\bx^2/\sigma^2-\sigma^2\bp^2\right]\,.
 \eeq
After lengthy, but straightforward algebra we thus obtain the source
Wigner density 
 \beq
 \label{n6}
   g(x,\bp) = {E_p\,\delta(x^0)\over (2\pi R\Delta)^3}\,
   \exp\left[-{(\bx - s L^2\bp)^2\over 2R^2}
   -{\bp^2\over 2\Delta^2}\right]\,,
 \eeq
where \cite{Cris98_2,fn1}
 \begin{mathletters}
 \label{n7}
 \bea
 \label{n7a}
   &&L^2 = {R_0 \Delta_0\over \Delta^2}\,,
 \\
 \label{n7b}
  &&R^2 \Delta^2 = R_\sigma^2\Delta^2 - s^2 R_0^2 \Delta_0^2
    \geq {1\over 4}\,,
 \\
 \label{n7c}
  &&R_\sigma^2 = R_0^2 + {\sigma^2\over 2}\,, \quad
    \Delta^2 = \Delta_0^2 + {1\over 2\sigma^2}\,.
 \eea
 \end{mathletters}
Eq.~(\ref{n6}) differs from (\ref{55}) only by a $\bp$-dependent
shift of the spatial coordinate $\bx$. The new length scale $L$
characterizes this $x$-$p$-correlation. Otherwise the two expressions
are completely identical, after redefining the physical parameters
$R,\Delta$ according to (\ref{n7}). The corresponding two-particle
exchange amplitude is given by 
 \beq
 \label{n8}
   \bar\rho(\bp_1,\bp_2) = 
   {\sqrt{E_1 E_2}\over (2\pi\Delta^2)^{3\over 2}}\,
   \exp\Bigl( - {\bK^2\over 2\Delta^2} 
              - {R^2\bq^2\over 2} - i s L^2 \bK{\cdot}\bq\Bigr).
 \eeq
This differs from (\ref{56}) only by the phase factor involving $L^2$.
Since $\bK{\cdot}\bq = (\bp_1^2-\bp_2^2)/2$, in the ring integrals
(\ref{21}) the phase factors involving the integration momenta cancel, 
leaving only those corresponding to the external momenta. Since  
the particle spectra are real and always involve products of such
phase factors with their complex conjugates ({\it cf.} Eqs.~(\ref{31})
and (\ref{37})), they completely drop out from the calculation of
higher-order multi-boson effects. The algebra of the previous
subsection thus goes through without modification; the $s$-dependence
of (\ref{n1}) which describes the strength of $x$-$p$-correlations in
the source has been completely absorbed into the redefined source 
parameters $R,\Delta$.  

\section{Results}
\label{sec5}

In this section we give a comprehensive and quantitative discussion of 
the behaviour of the correlation functions, the correlation strength,
and the HBT radius parameters, both for events with fixed multiplicity
and for multiplicity-averaged event ensembles. The results for fixed
multiplicity can be obtained in two different ways: either by using
Eqs.~(\ref{30})--(\ref{32}), or from the expression (\ref{38}) for
multiplicity-averaged events with the special multiplicity
distribution $p_m{=}\delta_{mn}$. We have checked that both paths lead
to identical results, but the second path allows for an easier presentation.

\subsection{Correlation functions: Bose-Einstein and residual correlations}
\label{sec5a}

We begin by rewriting (\ref{38}) as follows:
 \bea
 \label{62}
   &&C_2(\bp_1,\bp_2) = 
     {\sum_{i,j=1}^\infty h_{i+j}\, G_i(\bp_1,\bp_1)\, G_j(\bp_2,\bp_2)
      \over
      \sum_{i,j=1}^\infty h_i\, h_j\, G_i(\bp_1,\bp_1)\, G_j(\bp_2,\bp_2)}
 \nonumber\\
   &&\quad \times \left( 1 +
   {\sum_{i,j=1}^\infty h_{i+j}\, G_i(\bp_1,\bp_2)\, G_j(\bp_2,\bp_1)
    \over
    \sum_{i,j=1}^\infty h_{i+j}\, G_i(\bp_1,\bp_1)\, G_j(\bp_2,\bp_2)}
   \right)
 \nonumber\\
   && \quad \equiv C_2^{({\rm res})}(\bq,\bK) 
   \Bigl( 1 + R_2(\bq,\bK) \Bigr) \, .
 \eea
The second term in brackets, $R_2$, arises from Bose-Ein\-stein
symmetrization of the two measured momenta, $\bp_1$ and $\bp_2$. As 
can be seen from the discussion in Sec.~\ref{sec3c}, in particular 
Eq.~(\ref{52}), this is the term which contains, via a Fourier 
transform, information about the effective source size; to
extract this information is the goal of HBT interferometry. The 
expression inside the bracket shows manifestly that the source
is fully chaotic, as discussed at the end of Sec.~\ref{sec3d}: at
$\bq{=}0$ we have $R_2=1$, and both terms inside the bracket become 
equal.

The interesting feature of Eq.~(\ref{62}) is the prefactor 
$C_2^{({\rm res})}(\bq,\bK)$ which, following Zajc \cite{Zajc87,Zajc84}
who first noticed this effect, we call ``residual correlation''.
Equation (\ref{62}) clearly exhibits the physical origin of the 
residual correlations: they describe the result of the different 
combinatorics involved in creating background pairs from {\em mixed} 
events (the denominator of $C_2^{({\rm res})}$) and ``sibling pairs'' 
from the {\em same} event (the numerator of $C_2^{({\rm res})}$), 
without symmetrizing with respect to the two measured momenta $\bp_1$ 
and $\bp_2$. [Note that the measured momenta $\bp_1$ and $\bp_2$ are 
still symmetrized with respect to all other momenta of particles within 
the respective events; for this reason $C_2^{({\rm res})}$ can, 
unfortunately, {\em not} simply be interpreted (and measured separately) 
as the ratio of unlike-sign pairs from the same event divided by the 
like- or unlike-sign background pairs from mixed events: in that ratio
the numerator involves symmetrization only with respect to the momenta 
of those particles having the same charge as the measured ones, leading 
to different combinatorics.]
 
For sources of the type described in Sec.~\ref{sec2}, the resi\-dual 
correlations are absent if and only if the multipli\-ci\-ty distribution 
corresponds to the ``Poisson limit'' of Sec.~\ref{sec3c} (see 
Eq.~(\ref{53})) where $h_{i+j}{=}h_i h_j$. And only in this limit 
it is possible to isolate the information on the source geometry by
extracting from the Bose-Einstein correlator $R_2$ a simple ratio of 
particle spectra (see again (\ref{53})). Equation (\ref{62}) thus shows 
that the usual formalism for HBT interferometry must be significantly 
modified if one allows for arbitrary multiplicity distributions in a
situation where multi-boson symmetrization effects must be taken into 
account. This is true in particular for sources with a {\em fixed} 
pion multiplicity. While this need for a modified extraction procedure 
for the HBT radii was noticed and a specific suggestion for dilute 
sources was made in \cite{Letal00}, we present here a completely 
general approach to this problem.

The goal of the remainder of this section is to give a quantitative 
assessment of the importance of residual correlations and multi-boson 
effects on HBT radii, by exploring the full physical parameter range 
for the Gaussian sources discussed in Sec.~\ref{sec4} and for a variety 
of standard multiplicity distributions. This study reveals a number of 
unexpected new features of two-particle interferometry which have not 
been previously investigated. As we will show, they can be used in new 
ways to obtain information on the phase-space density of the source at 
freeze-out.

 \begin{figure}[ht]
 \epsfxsize=7cm \epsfysize=5.2cm
 \centerline{\epsfbox{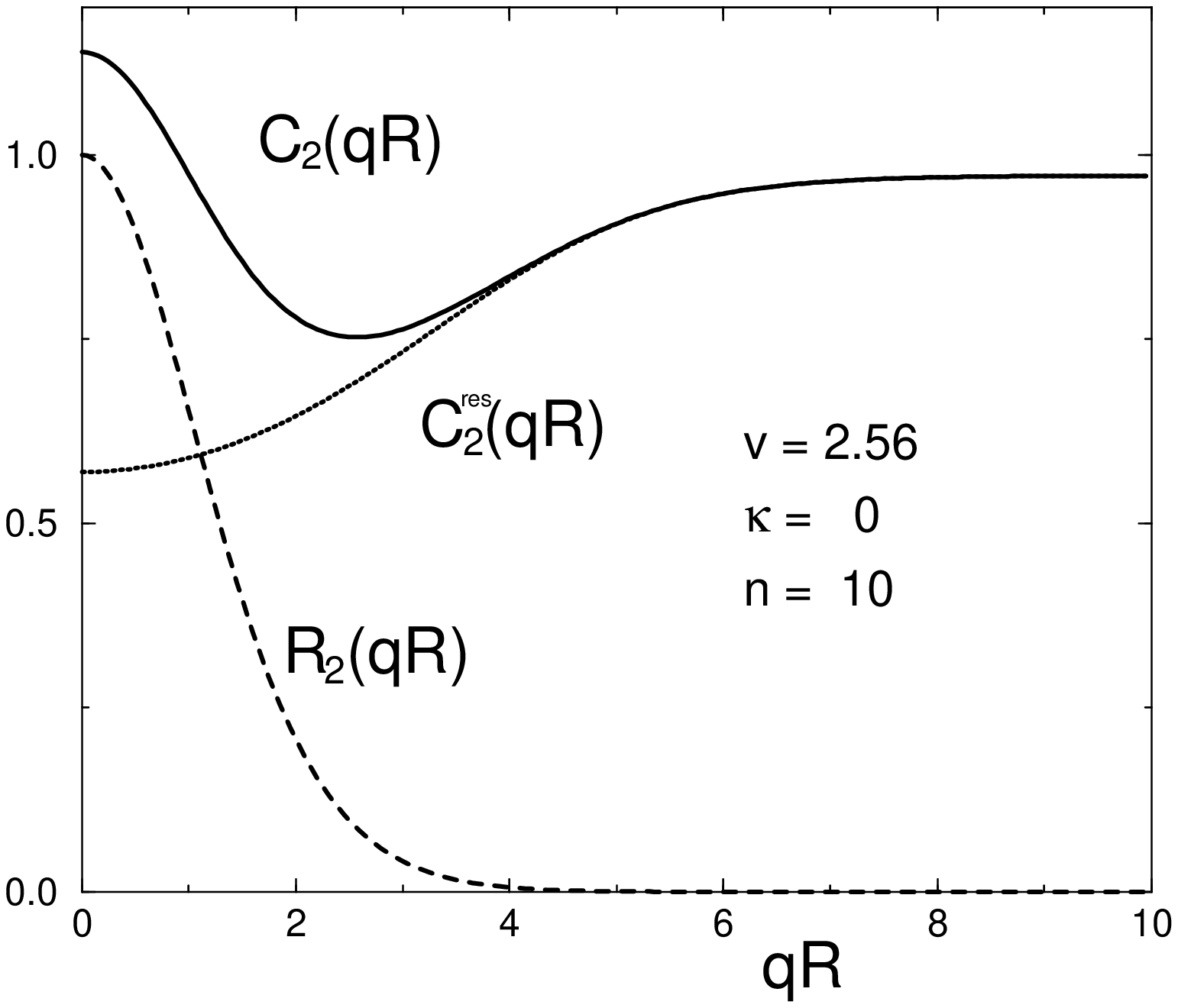}}
 \epsfxsize=7cm \epsfysize=5.2cm
 \centerline{\epsfbox{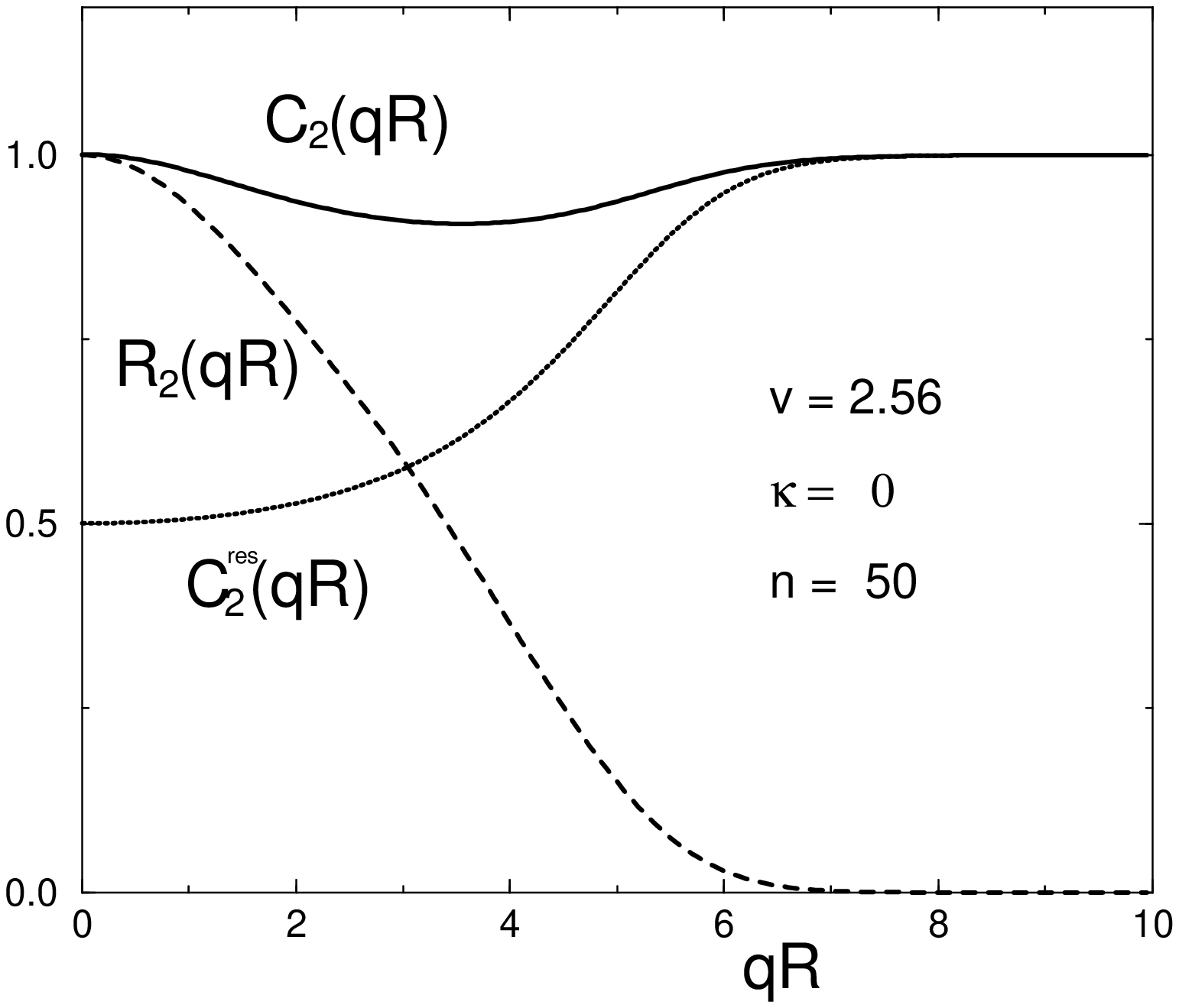}}
 \caption{Correlation function $C_2$ (solid), residual correlation
   $C_2^{({\rm res})}$ (dotted), and Bose-Einstein correlation $R_2$ 
   (dashed) for events of fixed multiplicity $n=10$ (top) and $n=50$ 
   (bottom), for vanishing pair momentum $\kappa$. (Note that for 
   $\kappa=0$ the correlation function is isotropic in $\bq$.) In 
   both plots the phase-space volume factor is set to $v=2.56$. 
 \label{F1}}
 \end{figure} 

It is convenient to introduce scaled momenta,
 \beq
 \label{63}
   \bbox{\kappa} = \bK/\Delta\,, \quad \tilde \bq = R\,\bq\,, 
 \eeq
and write $\kappa = \vert\bbox{\kappa}\vert,\, \tilde q = \vert \tilde
\bq\vert$. Pair momenta are thus measured in units of the width $\Delta$ 
of the source in momentum space, while relative momenta are scaled
by its width $R$ in coordinate space.

Explicit expressions for the correlation function (\ref{62}) for a
Gaussian source are given in Appendix~\ref{appa}. Fig.~\ref{F1}
shows two sample correlation functions for events with fixed 
multiplicities $n=10$ and $n=50$, respectively. Obviously, in both 
cases the correlation strength $\lambda$ (defined as the ratio of the
correlation function at $q=0$ and $q=\infty$ minus 1) is less than 
one. For the selected values of the pair momentum $\kappa$ and the 
phase-space volume $v$, the correlation strength actually vanishes 
at $n=50$. Fig.~\ref{F1} shows that this is caused by the residual 
correlations which become stronger with increasing phase-space 
density $n/v^3$, pulling the intercept at $q=0$ down. This effect 
is studied in more detail in Sec.~\ref{sec5c}.

For the case of fixed multiplicity studied in Figures \ref{F1} the 
residual correlations are {\em negative}, i.e. the correlation function 
approaches its asymptotic value at $q\to\infty$ {\em from below}. This
was found before in Fig.~6 of the last paper of Refs.~\cite{Zh97} 
(although not explicitly noted in the text), and the origin of the effect 
was correctly identified in \cite{Letal00} and independently in 
\cite{Scotto}. We want to emphasize, however, that the negative sign of 
the residual correlations is {\em not generic}: as will be seen in 
Sec.~\ref{sec5c}, it depends on the multiplicity distribution of the 
emitted particles, and a Bose-Einstein distribution, for example, 
yields {\em positive residual correlations}. 

\subsection{Multiplicity distributions}
\label{sec5b}

Before proceeding further, we specify in this subsection a few particular
forms for the measured pion multiplicity distribution. They will be used
below to illustrate our general results with specific examples. In particular,
we will show results for:
 \begin{mathletters}
 \label{64}
 \bea
 \label{64a}
   p_n &=& \delta_{nm}\,, \qquad
   ({\rm fixed\ multiplicity}\ m)
 \\
 \label{64b}
   p_n &=& {\la  n\ra^n \over n!}
   e^{-\la  n\ra}\,,
 \nonumber\\
   &&{\rm (Poisson\ distribution)}
 \\
 \label{64c}
   p(n,k) &=& {(n{+}k{-}1)! \over n!\, (k{-}1)!} 
   {\left({\la  n\ra\over k}\right)^n \over
    \left(1{+}{\la  n\ra\over k}\right)^{n+k}} \,,
 \nonumber\\
    && {\rm (negative\ binomial\ distribution)}
 \\
 \label{64d}
   p_n^{^{\rm BE}} &=&
   {\la  n\ra^n \over
    \left(1{+}\la  n\ra\right)^{n+1}}\,,
 \nonumber\\
   &&{\rm (Bose{-}Einstein\ distribution)}  
 \\
 \label{64e}
   p_\Gamma(n,k) &=& {1\over n\Gamma(k)} 
   \left( {kn\over\la  n\ra} \right)^k
   e^{-kn/\la  n\ra} \,,
 \nonumber\\
   && {\rm (Gamma\ distribution).}
 \eea
 \end{mathletters}
The Poisson distribution (\ref{64b}) is obtained from the negative 
binomial (NB) distribution (\ref{64c}) by setting $k\to\infty$; the
Bose-Einstein distribution (\ref{64d}) is obtained for $k{=}1$. For 
the Poisson distribution one has 
 \bea
 \label{65}
   \la  n(n{-}1)\ra &=& \la  n\ra^2 
   \quad \Longleftrightarrow \quad
   \la  n^2 \ra - \la  n\ra^2 = \la  n\ra
 \nonumber\\
   && \qquad {\rm (Poisson)}
 \eea
while the Bose-Einstein distribution gives a larger variance:
 \bea
 \label{66}
   \la  n(n{-}1)\ra &=& 2\, \la  n\ra^2 
   \quad \Longleftrightarrow \quad
   \la  n^2 \ra - \la  n\ra^2 = \la  n\ra
   \bigl(1+\la  n\ra\bigr)
 \nonumber\\
   && \quad {\rm (Bose{-}Einstein)}\,.
 \eea
For the case (\ref{64a}) of fixed multiplicity the variance vanishes, of 
course.

\subsection{Normalization and correlation strength}
\label{sec5c}

The Bose-Einstein correlation $R_2$ equals 1 at $q=0$ and vanishes at 
$q\to\infty$. The asymptotic normalization of the correlation function 
$C_2$ and the value of the correlation strength are thus controlled by 
the residual correlations $C_2^{({\rm res})}$. 

The normalization ${\cal N}$ does not depend on the pair momentum
$\kappa$; the results of Appendix~\ref{appa} give
 \begin{mathletters}
 \label{67}
 \bea
 \label{67a}
   {\cal N} &\equiv& \lim_{q\to\infty} C_2(\tilde\bq,\bbox{\kappa}) =
   {h_2\over(h_1)^2}\,,
 \\
 \label{67b}
     \lim_{v\to\infty}{\cal N} &=& 
     {\la  n(n{-}1)\ra \over \la  n\ra^2}
     = \left\{ \begin{array}{lr}
               2\quad{\rm Bose{-}Einstein\,,}\\
               1\quad{\rm Poisson\,,}
               \end{array}\right. 
 \\
 \label{67c}
   \lim_{v\to 1}{\cal N} &=& {1{-}p_0{-}p_1 \over (1{-}p_0)^2}
 \nonumber\\
   &=& \left\{ \begin{array}{ll}
              1 &\ \ {\rm Bose{-}Einstein\,,}\\
             {1-(\la  n\ra{+}1)e^{-\la  n \ra}
              \over (1-e^{-\la  n \ra})^2} &\ \ {\rm Poisson\,.}
             \end{array}\right. 
 \eea
 \end{mathletters}  
For fixed multiplicity $n$ this reduces to
 \begin{mathletters}
 \label{68}
 \bea
 \label{68a}
   {\cal N}^{(n)} &=&  {\omega(n)\,\omega(n-2) \over \omega^2(n-1)}\,,
 \\
 \label{68b}
   \lim_{v\to\infty}{\cal N}^{(n)} &=& {n(n-1)\over n^2}\,,
 \\
 \label{68c}
   \lim_{v\to 1}{\cal N}^{(n)} &=& 1\,.
 \eea
 \end{mathletters}  
The generic difference of ${\cal N}$ from unity can be clearly seen in the 
upper panel of Figure~\ref{F1}. 

For the correlation strength one finds, using results from 
Appendix~\ref{appa},
 \begin{mathletters}
 \label{69}
 \bea
 \label{69a}
   \lambda(\kappa) &=& 
   2\,{(h_1)^2\over h_2}\,{A(\kappa)\over B(\kappa)} - 1\,,
 \\
 \label{69b}
   \lim_{v\to\infty} \lambda(\kappa) &=& 1\, ,
 \\
 \label{69c}
   \lim_{v\to 1} \lambda(\kappa) &=& 
       {(1{-}p_0)^2\over 1{-}p_0{-}p_1}\, 
       {\la  n(n{-}1)\ra \over \la  n \ra^2} - 1
 \\
   &=& \left\{ \begin{array}{ll}
              1 &\ \ {\rm Bose{-}Einstein\,,}\\
              e^{-\la  n \ra} 
              {\la  n\ra - 1 + e^{-\la  n \ra}
              \over 1-(\la  n\ra{+}1) e^{-\la  n \ra}} 
              &\ \ {\rm Poisson\,.}
             \end{array}\right.
 \nonumber 
 \eea
 \end{mathletters}
For fixed multiplicity $n$ this reduces to
 \beq
 \label{70}
   \lim_{v\to\infty} \lambda^{(n)}(\kappa) = 1\,,\quad
   \lim_{v\to1} \lambda^{(n)}(\kappa) = -{1\over n}\,.
 \eeq
One sees that for fixed $n$ in the limit $v\to 1$ (i.e. for small 
sources which saturate the uncertainty relation) the residual 
correlations become so strong that the effective correlation strength
$\lambda$ becomes {\em negative}. 

 \begin{figure}[ht]
 \epsfxsize=7.5cm \epsfysize=5cm
 \centerline{\epsfbox{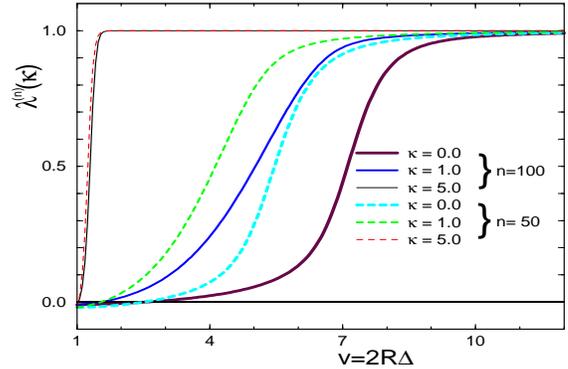}}
 \caption{Correlation strength $\lambda^{(n)}(\kappa)$ as a function 
  of the dimensionless phase-space volume $v=2R\Delta/\hbar$. Results 
  are shown for two fixed multiplicity values ($n=50$ (solid) and 100 
  (dashed)) and for three values of the reduced pair momentum 
  $\kappa=K/\Delta=0,\, 1$, and 5 (from right to left).
 \label{F2}}
\end{figure} 

Closer inspection of Eq.~(\ref{69c}) for the Poisson multiplicity 
distribution shows that it is not really the phase-space volume factor 
$v$, but the average phase-space density 
 \beq
 \label{71}
   d = {\la  n\ra\over v^3}
 \eeq
which controls the strength of multi-boson effects. For a source of
minimal phase-space volume $v{=}1$, the correlation strength $\lambda$ 
approaches zero as $\la  n\ra{\to}\infty$ (large phase-space 
density $d\to\infty$); this agrees with the expectation of 
strong multi-boson effects in this limit. On the other hand, in the 
opposite limit $\la  n\ra\to 0$ (small phase-space density
$d\to 0$) $\lambda$ approaches unity, the same value as for sources 
with large phase-space volumes $v\to\infty$. For strong multi-boson 
effects to arise, it is thus not sufficient that the phase-space 
{\em volume} factor $v$ is small, but the phase-space {\em density}
$d=\la  n\ra/v^3$ must be large.   

\begin{figure}[ht]
 \epsfxsize=7.5cm \epsfysize=5cm 
 \centerline{\epsfbox{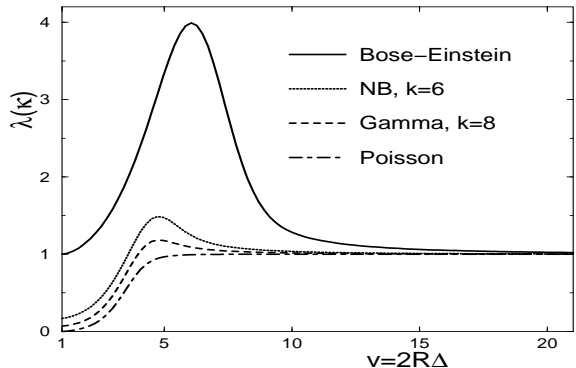}}
 \caption{Correlation strength $\lambda(\kappa)$ as a function of the 
  dimensionless phase-space volume $v=2R\Delta/\hbar$, for fixed pair 
  momentum $\kappa$=0.3. Results are shown for four different pion 
  multiplicity distributions (Poisson, Bose-Einstein, Gamma distribution 
  with $k$=8, and negative binomial (NB) distribution with $k$=6). In all 
  cases we fixed $\la  n\ra$=18.
 \label{F3}}
 \end{figure}

Let us now discuss some numerical results. In Figures \ref{F2} and 
\ref{F3} we show the correlation strength $\lambda$ as a function of 
the phase-space volume factor $v=2R\Delta/\hbar$. Whereas, according 
to Eqs.~(\ref{69}) and (\ref{70}), the correlation strength is 
independent of the scaled pair momentum $\kappa$ in the limits 
$v{\to}\infty$ and $v{\to}1$, Figure~\ref{F2} shows that in the 
intermediate range $1<v<\infty$ the correlation strength 
$\lambda^{(n)}$ develops a $\kappa$-dependence. At fixed phase-space 
volume $v$, the effective strength of the correlations diminishes for 
smaller $\kappa$ and/or larger multiplicity $n$. 

The dependence of the correlation strength on the multiplicity 
distribution is shown in Figure~\ref{F3}. For fixed $\kappa=0.3$ 
and constant mean multiplicity $\la  n\ra=18$, Fig.~\ref{F3} 
gives $\lambda$ as a function of $v$ for four different multiplicity 
distributions. One sees that for different multiplicity 
{\em distributions} the residual correlations manifest themselves 
quite differently. There is no universal scaling with the average 
phase-space density $d$ defined above; however, for typical momenta 
$\kappa\simeq 1$, significant residual correlation effects (i.e. 
noticeable deviations of $\lambda$ from its asymptotic value 
$\lim_{v\to\infty}\lambda{=}1$) are seen for phase-space densities 
above 0.2--0.3 particles per unit phase-space volume, $d\gtrsim 0.2-0.3$. 
At lower momenta $\kappa < 1$ similarly strong effects arise from even 
smaller values of $d$, whereas only very high phase-space densities 
lead to strong multi-boson effects at large pair momenta $\kappa\gg 1$. 
These conclusions agree with those of Lednicky {\it et al.} \cite{Letal00}. 

Note that the multiplicity distributions studied in Fig.~\ref{F3} 
refer to the {\em observed} pion multiplicity, after full inclusion of
multi-boson effects; the measured multiplicity distribution is kept 
fixed while varying the phase-space density. The presentation in
Fig.~\ref{F3} differs from the discussion of the ``Poisson limit''
in Sec.~\ref{sec3c} where the {\em input} distribution was
fixed to be Poissonian, and the measured multiplicity distribution 
$\bar p_n(v)$ of Eq.~(\ref{40}) then depended on $v$. The distribution 
$\bar p_n(v)$ corresponds to a constant correlation strength 
$\lambda{=}1$ and would in Fig.~\ref{F3} interpolate as a constant 
horizontal line between the Poisson distribution at $v{\to}\infty$ 
and the Bose-Einstein distribution at $v{\to}1$. This must be taken 
into account when comparing with results from the ``Poisson limit'' 
like those discussed in \cite{ZC98,ZSH98,Letal00}. 

\subsection{HBT radii}
\label{sec5d}

The information on the space-time structure of the source is contained
in the Bose-Einstein correlation function $R_2(\bq,\bK)$ of (\ref{62}).
The discussion of the ``Poisson limit'' in Sec.~\ref{sec3c} ({\em cf.}
Eq.~(\ref{53})) showed that this information is ``contaminated'': the 
two-particle exchange term
$\sum h_{i{+}j}\,G_i(\bp_1,\bp_2)\,G_j(\bp_2,\bp_1)$
is normalized by the corresponding ``direct term'' evaluated at the
observed momenta $\bp_1,\bp_2$ (i.e. by 
$\sum h_{i{+}j}\,G_i(\bp_1,\bp_1)\,G_j(\bp_2,\bp_2)$) instead of the
same expression evaluated at the average momentum $\bK$ (i.e.
$\sum h_{i{+}j}\,G_i(\bK,\bK)\,G_j(\bK,\bK)$). Only in the latter case
would $R_2(\bq,\bK)$ allow for a ``clean'' extraction of the space-time
structure of the source. For heavy-ion collisions where the sources are 
large it is known \cite{CSH95,P97} that (at least in the ``Poisson 
limit'') these contamination effects from inadequate normalization are 
negligible. For small sources, like those created in $e^+e^-$ collisions, 
they may, however, seriously affect the extraction of the source radius. 

In the ``Poisson limit'' Eq.~(\ref{53}) shows that this type of 
contamination can be avoided by first extracting from $R_2$ a 
(directly measurable) ratio of single particle spectra. For a 
dilute Gaussian source of the type (\ref{55}) (i.e. in the limit 
$v\to\infty$) this ratio is given by
 \beq
 \label{72}
   {[N_1(\bK)]^2\over N_1(\bp_1)N_1(\bp_2)} = 
   \exp\left( {q^2\,\over 4 \Delta^2}\right) = 
   \exp\left( {\tilde q^2\over v^2}\right)\,.
 \eeq
For arbitrary multiplicity distributions the residual correlations 
prohibit such a simple procedure (the denominator of $R_2$ cannot in 
general be written as a product of single particle spectra), and the 
contamination of the space-time information about the source cannot 
be avoided. The results presented in this subsection allow, however, 
to estimate the relative magnitude of this contamination effect. 

As usual, we parametrize the correlation function $R_2(\bq,\bK)$ by a 
Gaussian in $\bq$ with $\bK$-dependent width parameters (HBT radii). 
The spherical symmetry of the Gaussian source (\ref{55}) ensures that  
the HBT radii depend only on the modulus $\kappa$ of the scaled pair 
momentum (see Appendix~\ref{appb}). However, since the denominator of
$R_2$ involves the two measured momenta $\bp_1$ and $\bp_2$ (see 
(\ref{62}) and (\ref{A4})), the correlation function $R_2$ depends 
separately on the components $q_\parallel$ and $q_\perp$ of $\bq$ 
which are parallel and orthogonal to $\bK$. This is an artifact of
the above-mentioned ``contamination effect''; the difference of the
width parameters in the $q_\parallel$ and $q_\perp$ directions thus 
is a quantitative measure for this undesired contamination.

We thus write
 \bea
 \label{73}
   &&R_2(\bq,\bK) = \exp\left[- q_\perp^2\,R_\perp^2(K) 
                              - q_\parallel^2\,R_\parallel^2(K)\right]
 \, ,  
 \\
 \label{74}
   && q_\perp = q\,\sin\theta\,, \quad
      q_\parallel = q\,\cos\theta\, ,\quad
      \theta = \angle(\bq,\bK)\,.
 \eea
The HBT radii are then obtained by separating the constant and 
$\theta$-dependent terms in
 \bea
 \label{75}
    - {d\over dq^2}\,R_2(\bq,\bK) \bigg\vert_{q{=}0}
    &=& R_\perp^2(K) 
 \\
    && + \cos^2\theta \left(R_\parallel^2(K)-R_\perp^2(K)\right)\,.
 \nonumber
 \eea
The corresponding explicit expressions are given in Appendix~\ref{appb},
Eqs.~(\ref{B1}) and (\ref{B2}). Before proceeding further it is useful 
to establish a baseline in the absence of multi-boson effects, by keeping
in (\ref{B1}) and (\ref{B2}) only the lowest order terms $i{=}j{=}1$.
Denoting these reference values by a tilde we find
 \beq
 \label{76}
   \tilde R_\perp^2 = \tilde R_\parallel^2 = 
   R^2 \left(1-{1\over v^2}\right)
 \eeq
independent of the pair momentum $\kappa$. The HBT radii are seen to be
smaller than the Gaussian input radius $R$ by a correction $\sim 1/v^2$;
this is the ``contamination'' from the ratio (\ref{72}) of single 
particle spectra mentioned above. In the following we will discuss
multi-boson effects on the HBT radii (including additional 
multi-boson-induced ``contamination'' effects) as deviations from 
the baseline (\ref{76}). To this end we define the dimensionless ratios
 \beq
 \label{77}
   r_1 = {R_\perp^2\over R^2 \left(1-{1\over v^2}\right)}\,,
   \quad
   r_2 = {R_\parallel^2 - R_\perp^2 \over R^2}\,.
 \eeq
Like the HBT radii themselves, these ratios depend on $v$, $\kappa$ and 
$\la  n\ra$ via multi-boson symmetrization effects; in the absence
of such effects (in particular in the limit $v\to\infty$) they approach
$r_1 = 1$ and $r_2 = 0$. In particular, $r_2$ provides a relative measure
for multi-boson-induced spectral contamination effects.

 \begin{figure}[ht]
 \epsfxsize=7cm \epsfysize=5cm
 \centerline{\epsfbox{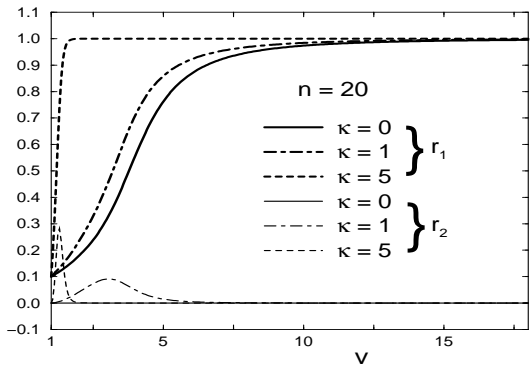}}
 \caption{The ratios $r_1$ (thick lines) and $r_2$ (thin lines) as
   functions of $v=2R\Delta/\hbar$, for fixed multiplicity $n=20$ and
   for three values of the scaled pair momentum $\kappa=0,1,5$. $r_2$
   vanishes identically for $\kappa=0$ (see Eq.~(\ref{B2})).
 \label{F4}}
 \end{figure} 

Analytic expressions for $r_1$ and $r_2$ in the quantum saturation 
limit $v\to 1$ (in which the pions in the source all occupy the same
quantum state and form a Bose condensate) are given in Eqs.~(\ref{B3}).
Note that in this limit both $R_\perp$ and $R_\parallel$ vanish 
{\em even in the absence of multi-boson effects}, due to the factor
$(1-{1\over v^2})$ in (\ref{76}). This leads to a flat Bose-Einstein 
correlation function $1+R_2$. In \cite{ZC98} this flatness of the 
correlation function was incorrectly attributed to the onset of phase 
coherence, resulting in an absence of Bose-Einstein correlations
at low values of $q$. Lednicky {\it et al.} \cite{Letal00} clarified 
this error by pointing out that even in this limit the Bose-Einstein 
correlations are still present (the exchange term $R_2=1$ is as large 
as the direct term), and that the misunderstanding in \cite{ZC98}
resulted from an incomplete analysis of the normalization of the
correlation function in the Bose-Einstein condensation limit. The 
latter was also discussed in \cite{ZSH98}.

 \begin{figure}[ht]
 \epsfxsize=7cm \epsfysize=5cm
 \centerline{\epsfbox{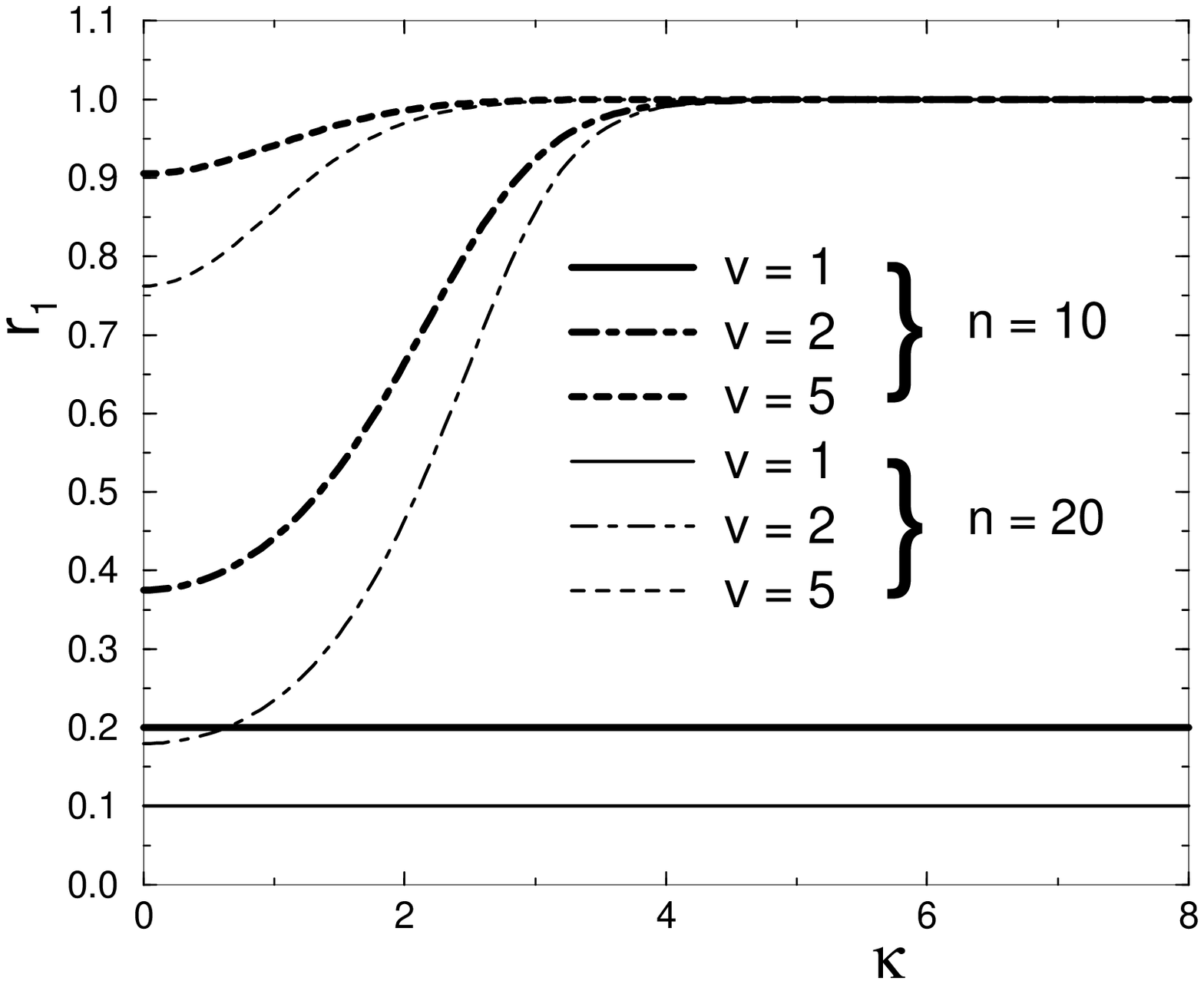}}
 \epsfxsize=7cm \epsfysize=5cm
 \centerline{\epsfbox{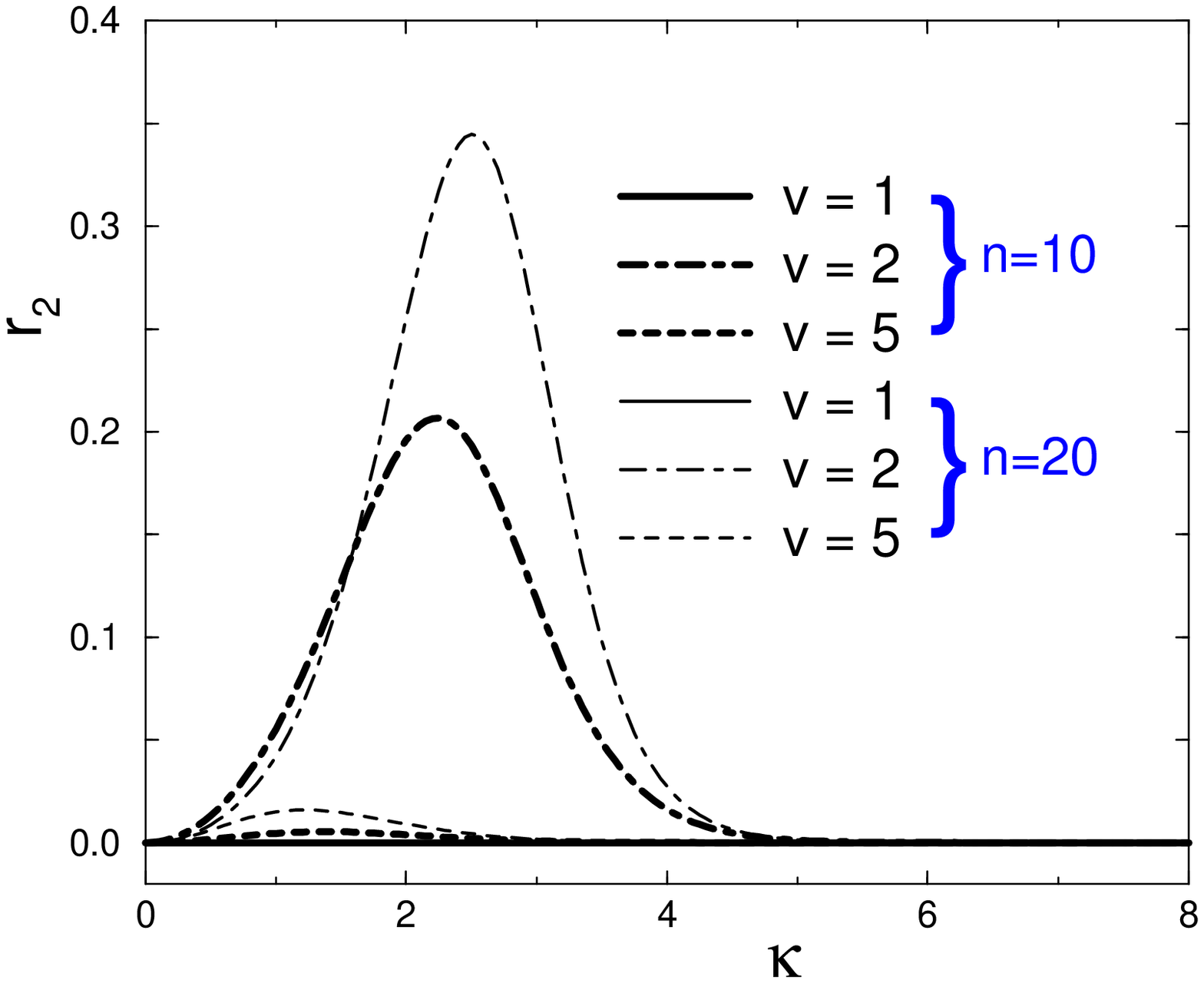}}
 \caption{Top: The ratio $r_1$ as a function of the scaled pair 
   momentum $\kappa$, for two fixed multiplicities ($n=10$ (thick 
   lines) and $n=20$ (thin lines)) and three values for the
   phase-space volume factor $v=1,2,5$. For $v=1$, $r_1$ is 
   independent of $\kappa$ and less than 1; but for $v=1+\epsilon$, 
   $r_1$ will eventually again approach 1 at $\kappa\to\infty$. 
   Bottom: The same for the ratio $r_2$; it vanishes for $v=1$.
 \label{F5}}
 \end{figure} 

Eqs.~(\ref{B3}) show that for all considered multiplicity distributions 
$\lim_{v\to 1} r_1 \leq 1$. The numerical results below show that the 
same holds true for all values of $v$. In other words, multi-boson 
effects always {\em reduce} the HBT radii: $R_\perp^2 \leq R^2 
 (1-{1\over v^2})$. For $v>1$ this is largely the consequence of 
Bose-clustering in the source ($R_n^2 = a_n R^2 \leq R$, see (\ref{57}))
and confirms similar findings in many earlier papers. However, for $v=1$ 
the source is already as small as allowed by the uncertainty relation 
and thus cannot be compressed any further by Bose-clustering. In this 
case a reduced value $r_1 < 1$ must be blamed entirely on multi-boson 
contributions to the spectral contamination effect from the product of 
single particle spectra in the denominator of the Bose-Einstein 
correlator $R_2$. Eqs.~(\ref{B3}) show that these contributions disappear 
for vanishing phase-space density $d = \la  n\ra/v^3\to 0$, even in 
the limit $v\to 1$:
 \beq
 \label{78}
   \lim_{\la  n\ra \to 0} \lim_{v\to 1} r_1 = 1\,.
 \eeq
(For fixed multiplicity the lowest allowed value is $n=2$ for which
$r_1$ satisfies the same rule.) However, for large phase-space
densities $d\to\infty$ they are large and the ratio $r_1$ vanishes:
 \beq
 \label{79}
   \lim_{\la  n\ra\to\infty} \lim_{v\to 1} r_1 =
   \lim_{\la  n\ra\to\infty} \lim_{v\to 1} 
   {R_\perp^2 \over R^2 \left(1-{1\over v^2}\right)} = 0\,.
 \eeq 
%
 \begin{figure}[ht]
 \epsfxsize=7cm \epsfysize=5cm
 \centerline{\epsfbox{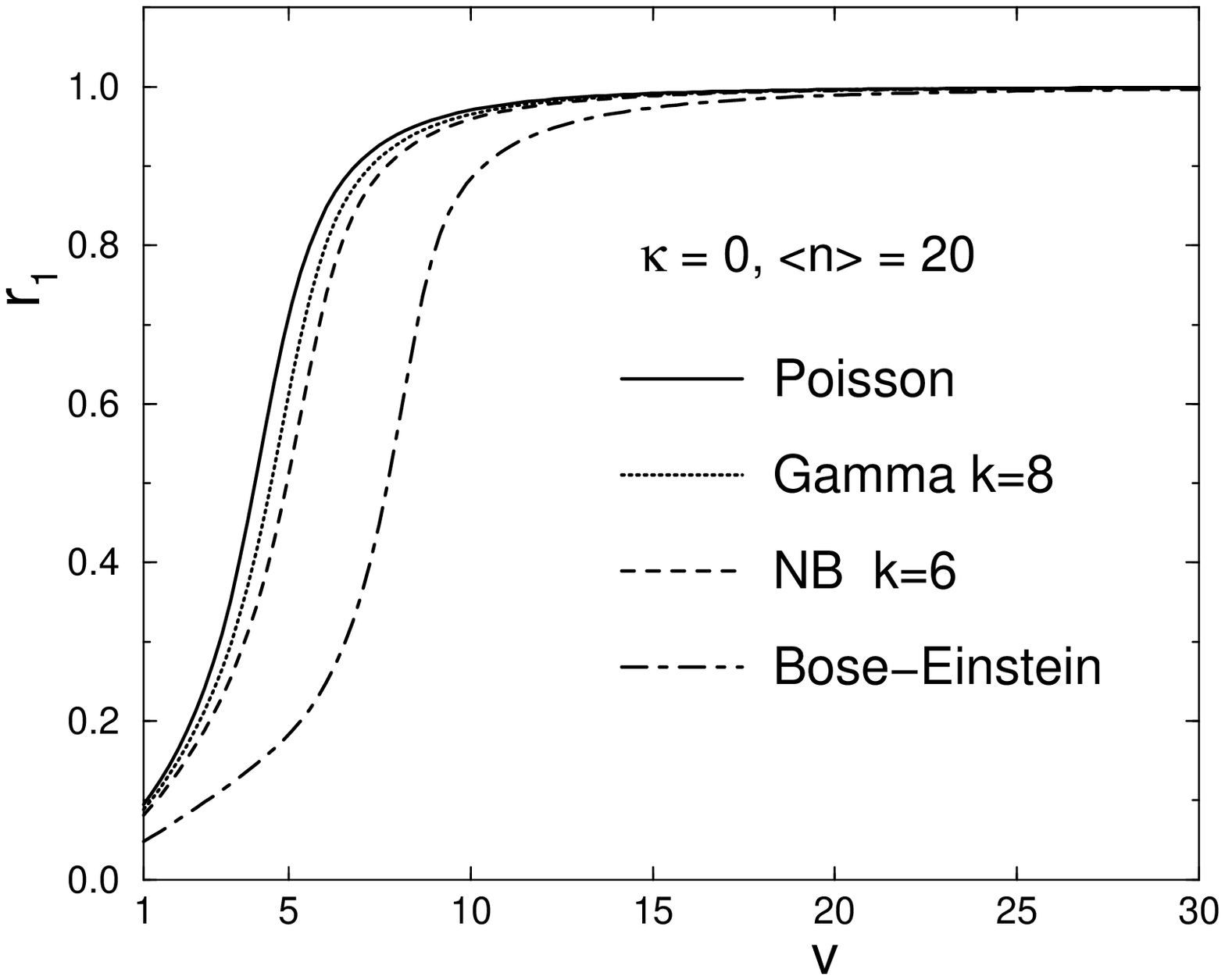}}
 \epsfxsize=7cm \epsfysize=5cm
 \centerline{\epsfbox{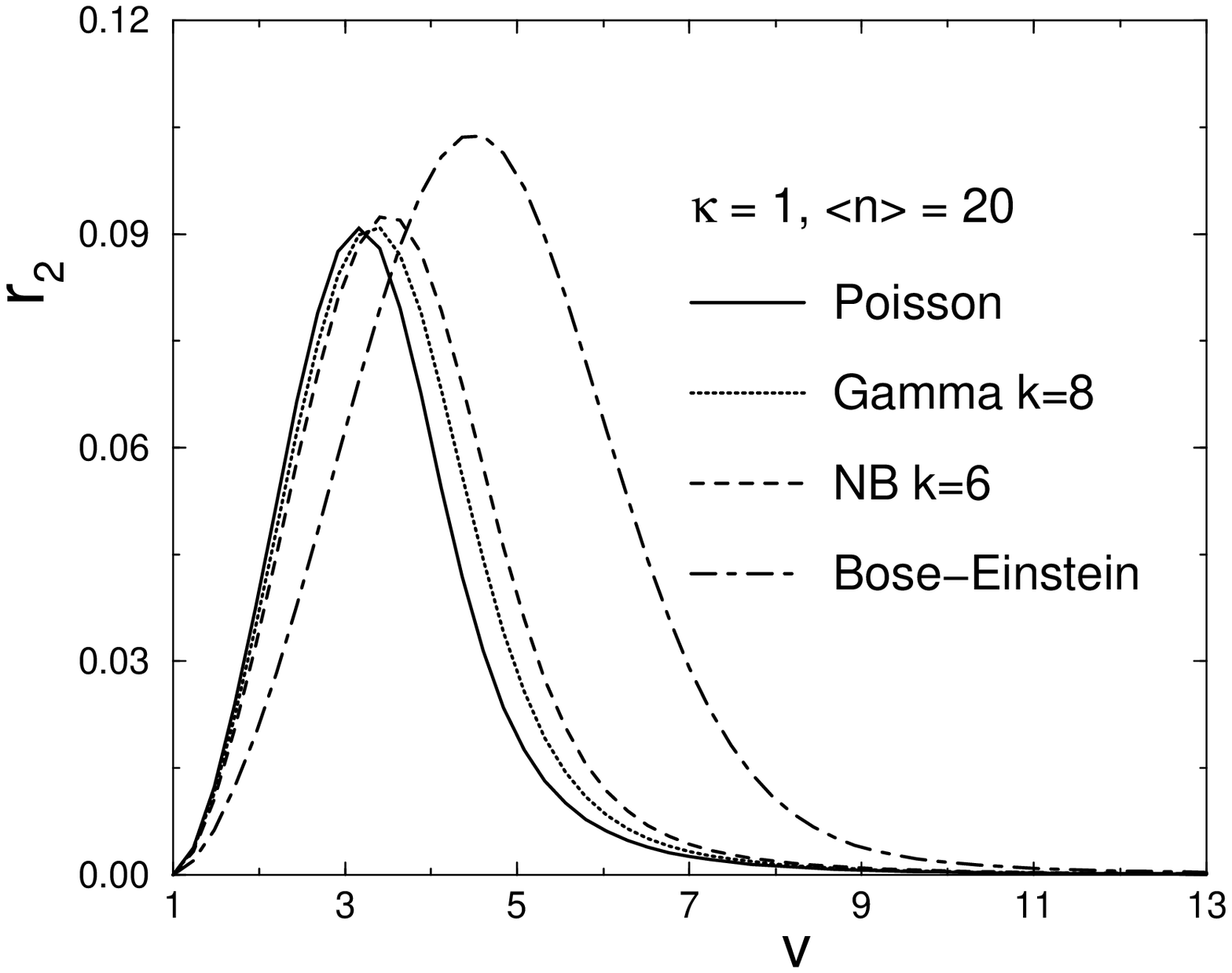}}
 \caption{Top: The ratio $r_1$ as a function of $v$, for vanishing
   pair momentum and fixed average multiplicity $\la  n\ra=20$,
   for four different multiplicity distributions. Bottom: The same for
   the ratio $r_2$, but for $\kappa{=}1$ (since $r_2{=}0$ for
   $\kappa{=}0$). 
 \label{F6}}
 \end{figure} 
%
\noindent
These last results can be generalized to arbitrary values of $v$ by
rephrasing them in terms of $d$: 
 \begin{mathletters}
 \label{a80}
 \bea
 \label{a80a}
   \lim_{d \to 0} r_1 &=& 1\,,
 \\
 \label{a80b}
   \lim_{d \to 0} r_2 &=& 0\,,
 \\
 \label{a80c}
   \lim_{d \to \infty} r_1 &=& \lim_{d \to \infty} r_2 = 0\,.
 \eea
 \end{mathletters}
For the limit $d\to \infty$ the proof relies on the fact that, 
at fixed $v$, this requires $\la n\ra\to\infty$; therefore only 
the asymptotic behavior of (\ref{57}) for large $n$ is needed, in 
particular 
 \beq
 \label{a81}
    \lim_{n\to\infty} a_n = {1\over v}\,.
 \eeq
(Incidentally, this implies $\lim_{n\to\infty} R_n\Delta_n =
 {R\Delta\over v} = {1\over 2}$: in the limit $n\to\infty$ the 
effective source corresponding to $G_n(\bp_1,\bp_2)$ saturates the 
uncertainty relation.) Equations (\ref{a80c}) then follow easily 
from (\ref{B1}) and (\ref{B2}).  

\begin{figure}[ht]
 \epsfxsize=7cm \epsfysize=5.2cm
 \centerline{\epsfbox{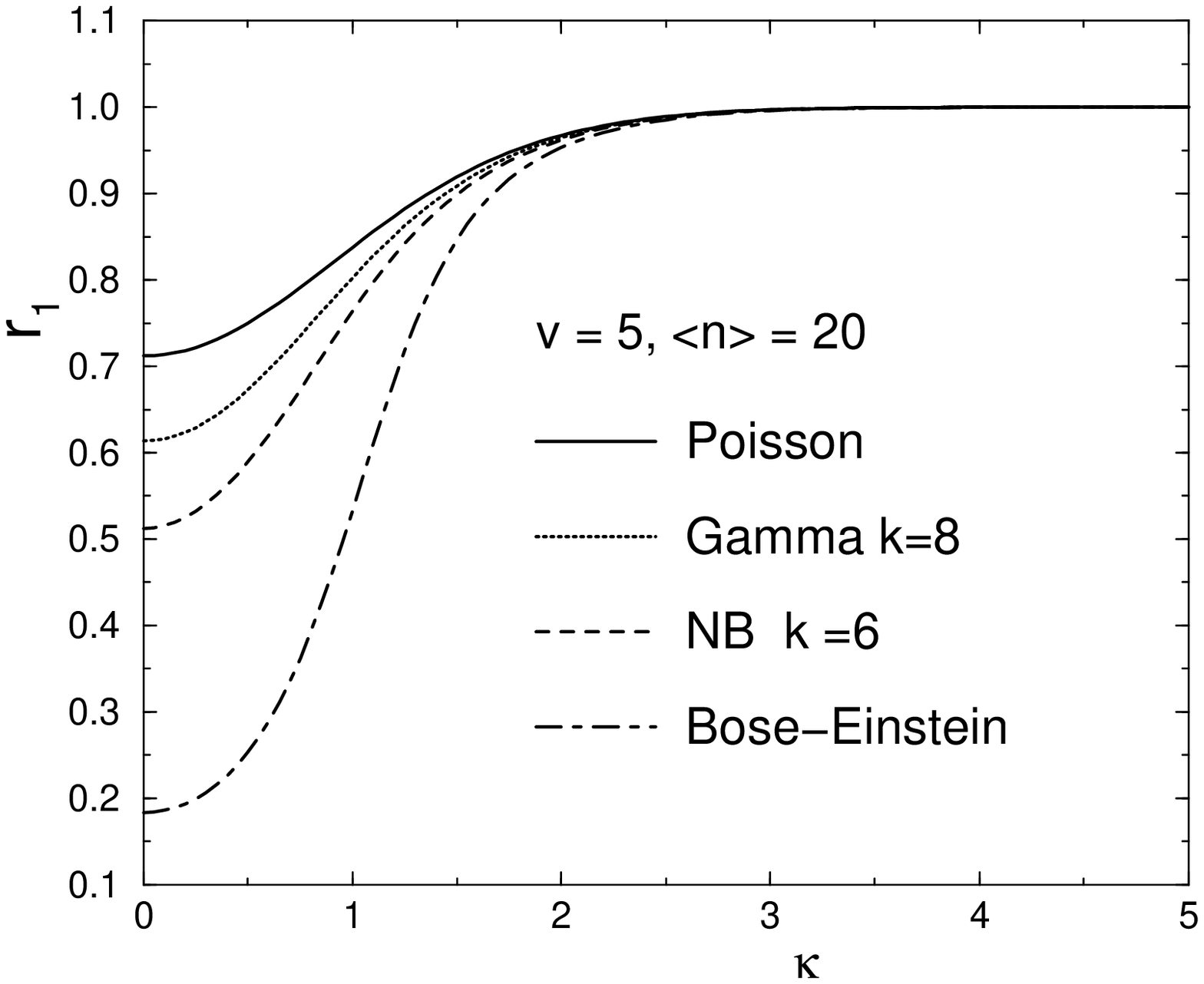}}
 \epsfxsize=7.15cm \epsfysize=5.2cm
 \centerline{\epsfbox{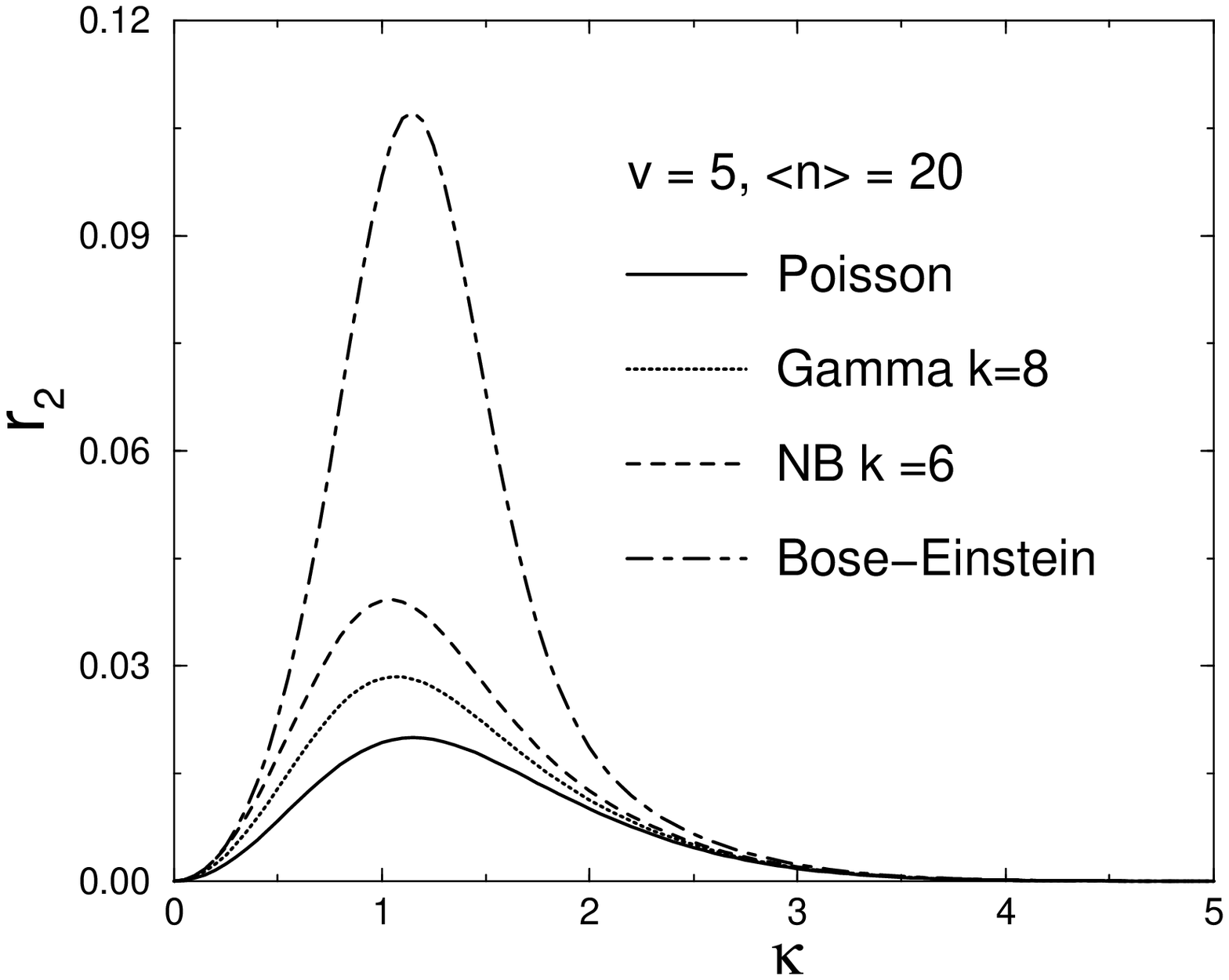}}
 \caption{Top: The ratio $r_1$ as a function of $\kappa$, for a fixed
   phase-space volume factor $v=5$ and fixed average multiplicity 
   $\la  n\ra=20$, for four different multiplicity distributions.
   Bottom: The same for the ratio $r_2$.
 \label{F7}}
 \end{figure}

Figures~\ref{F4}--\ref{F7} show numerical results for $r_1$ and $r_2$
for a variety of multiplicity distributions as functions of the 
phase-space volume factor $v$ and the scaled pair momentum $\kappa$.
In Figures \ref{F4} and \ref{F5} we study events with fixed 
multiplicity ($n=10$ and $n=20$). Multi-boson symmetrization effects
on the HBT radius $R_\perp$ and their contributions to the
difference $R_\parallel^2-R_\perp^2$ (which is entirely due to the
above-mentioned ``spectral contamination effect'') are seen to increase
as the phase-space density $d$ increases and the scaled pair momentum
$\kappa$ decreases. On the other hand, Fig.~\ref{F5} shows that for 
large pair momenta $\kappa$ the multi-boson effects vanish, irrespective 
of the value of $v$. As noted in \cite{Letal00} this reflects the 
exponential decrease of the momentum-space density $\sim \exp(-\kappa^2/2)$ 
at large particle momenta. This also explains why increasing $\kappa$ 
compresses the curves in Fig.~\ref{F4} to the left.
 
Note that the multi-boson effects set in rather suddenly. (As discussed 
in Appendix~\ref{appb}, the vanishing of $r_2$ in the limits $v\to 1$ 
and/or $\kappa\to 0$ is due to its definition and does not imply the 
disappearance of multi-boson effects in these limits.) Unfortunately, 
we have not been able to find a simple scaling law for the threshold at 
which multi-boson effects on the HBT radii set in and which properly 
describes the observed qualitative trends. 

In Figures \ref{F6} and \ref{F7} we study the dependence of
multi-boson symmetrization effects on the measured pion multiplicity 
distribution. While the dependences on $v$ (Fig.~\ref{F6}) and on
$\kappa$ (Fig.~\ref{F7}) are qualitatively similar to the case of
fixed multiplicity, on a quantitative level one observes quite
significant variations for different multiplicity distributions at 
fixed mean multiplicity $\la  n\ra$. This demonstrates that there is
no simple scaling of multi-boson effects with the average phase-space
density $d=\la  n\ra/v^3$. Of the four distributions considered here, 
the Bose-Einstein multiplicity distribution gives by far the strongest 
multi-particle symmetrization effects on the HBT radii: it is less 
effective in suppressing high-multiplicity contributions than the
other considered distributions. 

\subsection{The range of the residual correlations}
\label{sec5e}

Looking at Figure\,\ref{F1} and Appendix\,\ref{appa} one sees that the 
residual correlation $C_2^{({\rm res})}(\bq,\bK)$ can also be 
pa\-ra\-me\-trized by a Gaussian:
 \beq
 \label{80}
   C_2^{({\rm res})}(\bq,\bK) = {\cal N} 
   \left[ 1 + {\textstyle{\lambda(\kappa){-}1\over 2}}\,
   e^{-q_\perp^2\,r_\perp^2(\kappa)
      -q_\parallel^2\,r_\parallel^2(\kappa)}\right].
 \eeq
The two newly introduced radius parameters $r_\perp(\kappa)$,
$r_\parallel(\kappa)$ can be interpreted as the ``range'' of the 
residual correlations in coordinate space. A Gaussian parametrization
of the full correlation function $C_2(\bq,\bK)$ in (\ref{62}) is
obtained by combining (\ref{80}) with (\ref{73}). 

From the ansatz (\ref{80}) one obtains the ``residual correlation radii''
by separating the constant and $\theta$-dependent terms in
 \bea
 \label{81}
    &&{r_\perp^2(\kappa)\over R^2} + 
      \cos^2\theta {r_\parallel^2(\kappa)-r_\perp^2(\kappa)\over R^2}
 \nonumber\\
    && = {B(\kappa)\, \over {\cal N}\,B(\kappa)-A(\kappa)}\,
         {d\over d\tilde q^2}\,C_2^{({\rm res})}(\tilde\bq,\bbox{\kappa}) 
         \bigg\vert_{\tilde q{=}0}\,,
 \eea
where we used (\ref{67a}) and (\ref{69a}) to rewrite the coefficient
in front of the $\tilde q^2$-derivative. The explicit expressions are 
given in Appendix~\ref{appb}, Eqs.~(\ref{B4})--(\ref{B9}). Using 
the asymptotic expressions (\ref{A9}) we find from (\ref{B4}) and
(\ref{B5})
 \beq
 \label{82}
   \lim_{v\to\infty} r_\perp^2(\kappa) =
   \lim_{v\to\infty} r_\parallel^2(\kappa) = 0\,;
 \eeq
for dilute sources the residual correlations disappear, as expected. 
In the opposite limit $v\to 1$ the two ``residual correlation radii'' 
are nonzero, but again equal to each other:
 \beq
 \label{83}
   \lim_{v\to 1} \bigl(r_\perp^2(\kappa) - r_\parallel^2(\kappa)\bigr) 
   = 0\,.
 \eeq
What matters in practice is the ratio of the ``residual correlation
radii'' to the HBT radii, $r_\perp/R_\perp$. For fixed multiplicity $n$,
this ratio can be calculated analytically in the limit $v\to 1$:
 \beq
 \label{84}
   \lim_{v\to 1} {r_\perp^2(\kappa)\over R_\perp^2(\kappa)} =
   {1\over 4}\, {n-1\over n+1}  \quad
   ({\rm fixed}\ n).
 \eeq
For $n\to\infty$ this approaches the value ${1\over 4}$ from below; for 
intermediate values $1 < v < \infty$ the ratio is between 0 and 
${1\over 4}$. The same is true for the Poisson multiplicity distribution 
for which (\ref{B9}) gives in the limit of large $\la  n\ra$
 \beq
 \label{85}
   \lim_{v\to 1} 
   {r_\perp^2(\kappa)\over R_\perp^2(\kappa)} =
   {1\over 4}\, {\la  n\ra -2 \over \la  n\ra -1} + 
   {\cal O}\bigl(e^{-\la  n\ra}\bigr)  \quad
   ({\rm Poisson}).
 \eeq
From these analytical results it seems that the ``range'' of the 
residual correlations in coordinate space is always less than half 
the effective source size as measured by the HBT radii. Below we 
confirm this numerically. Thus, with two well-separated length 
scales involved, strong multi-boson effects should be clearly
recognizable in the 2-particle correlation function, as exemplified in 
Fig.~\ref{F1}. 

 \begin{figure}[ht]
 \epsfxsize=7cm \epsfysize=5.2cm
 \centerline{\epsfbox{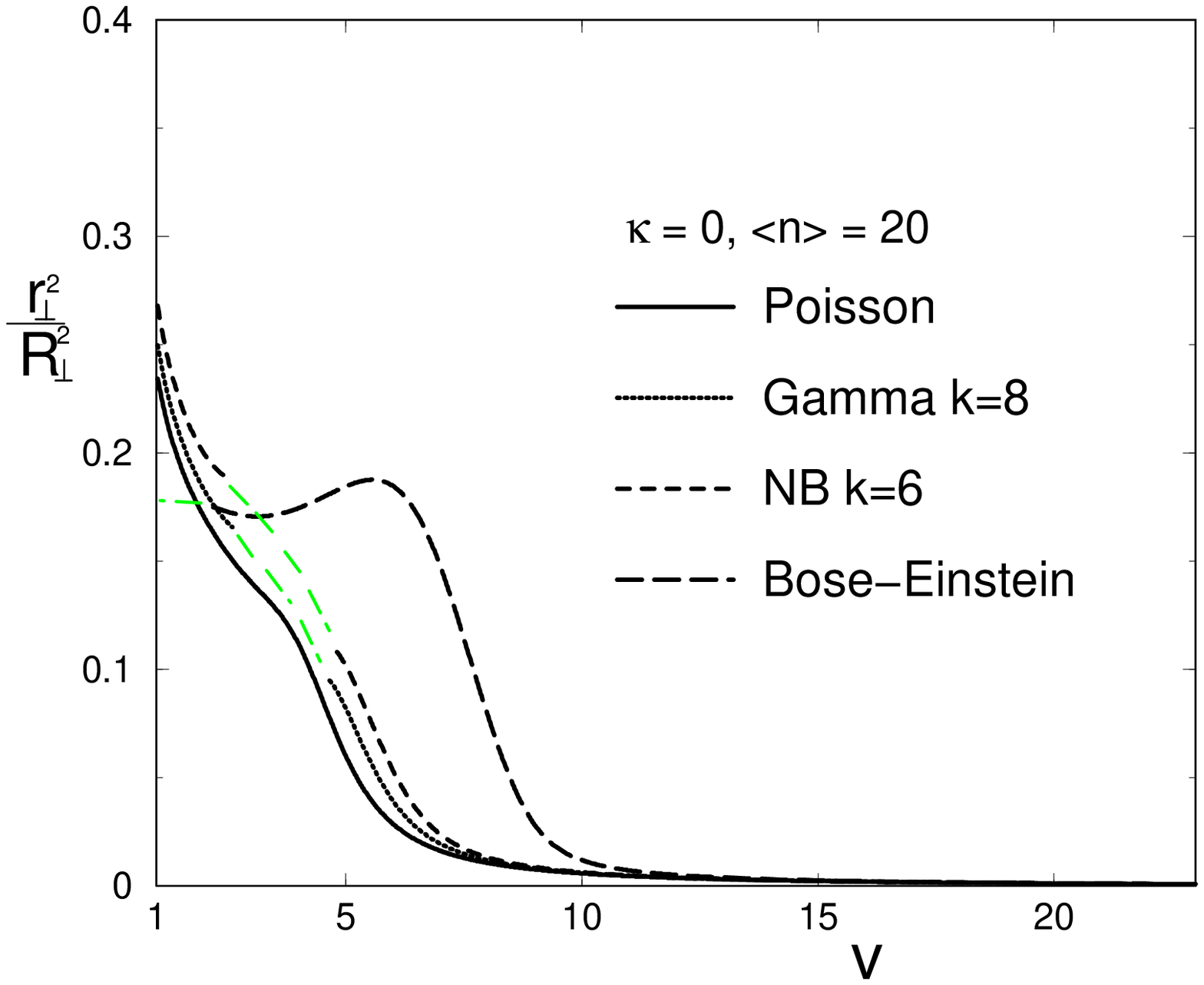}}
 \epsfxsize=7.15cm \epsfysize=5.2cm
 \centerline{\epsfbox{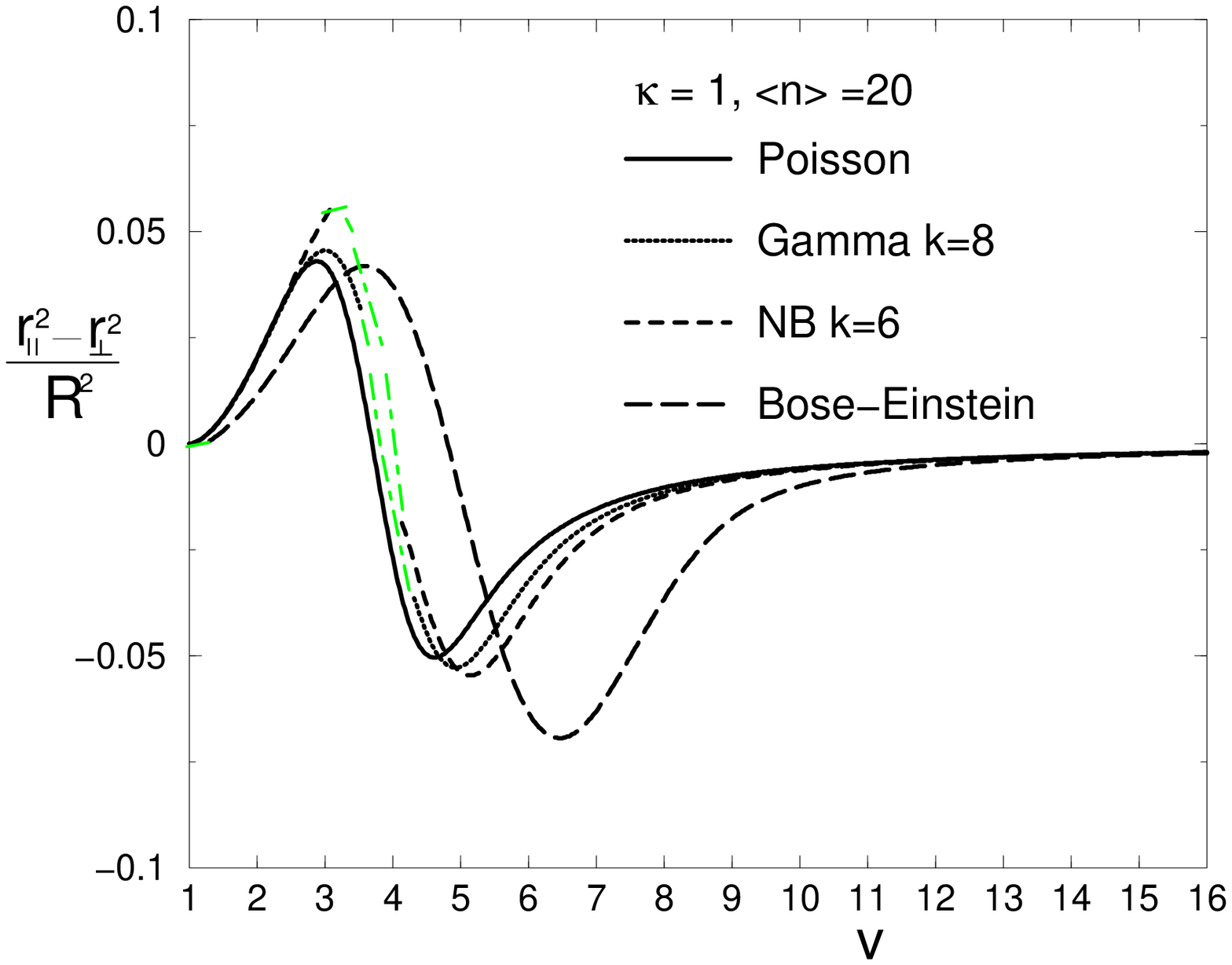}}
 \caption{Top: $v$-dependence of the ratio $r_\perp^2/R_\perp^2$ at 
   vanishing pair momentum $\kappa{\,=\,}0$, for a variety of 
   multiplicity distributions, keeping the average multiplicity 
   fixed at $\la n\ra{\,=\,}20$. Bottom: The same for 
   $(r_\parallel^2-r_\perp^2)/R^2$ at $\kappa{\,=\,}1$.
   The grey dashed lines indicate an interpolation as described 
   in the text. 
 \label{F8}}
 \end{figure} 

In Fig.~\ref{F8} we show the ratios $r_\perp^2/R_\perp^2$ and 
$(r_\parallel^2-r_\perp^2)/R^2$ as functions of $v$, for the same 
values of $\kappa$ and $\la n\ra$ as in Fig.~\ref{F6}. One sees that
indeed always $r_\perp \lesssim \half R_\perp$, and that the ratio
of these two parameters is largest at small values of $v$. The
ratio $(r_\parallel^2-r_\perp^2)/R^2$ alternates sign, but its
absolute value always stays below about 5\% and also below the
ratio $r_2$ of (\ref{77}) (see Fig.~\ref{F6}). 

We note that the range parameters $r_\perp^2(\kappa)$ and 
$r_\parallel^2(\kappa)$ become ill-defined when the correlation
strength $\lambda(\kappa)$ passes through 1 (see Eq.~(\ref{80}) and 
Fig.~\ref{F3}). We checked that at this point the function
$C_2^{\rm res}(\bq)$ is approximately, but not quite exactly 
constant. The remaining, albeit very weak $\bq$-dependence cannot
be properly absorbed by the {\em  Ansatz} (\ref{80}), resulting in a
divergence of the expression (\ref{81}). Since this is an artefact
of the {\em Ansatz} (\ref{80}) and does not reflect any really singular
behaviour of the residual correlations, we smoothed this singularity
in Fig.~\ref{F8} by interpolating by hand (grey dashed lines) across 
the point $\lambda(\kappa){\,=\,}1$.

\section{Experimental consequences}
\label{sec6}

Figure~\ref{F1} shows that, for fixed $n$, the residual correlations
reduce the correlation function while the Bose-Einstein correlations
increase it. The interplay of these effects leads to a minimum of the
correlation function at intermediate values of $q$. This is different 
from Coulomb or strong final state interactions (FSI) which are 
concentrated much more strongly near $q=0$ (and which in practice one
can correct for \cite{reviews}).  A minimum at nonzero $q$, after FSI 
correction, thus points to the presence of resi\-dual correlations arising 
from multi-boson symmetrization effects. In the limit $v\to\infty$, i.e. 
for vanishing phase-space density of the source, the residual 
correlations disappear, and so does the minimum in the correlation 
function. The minimum also disappears in the opposite limit, $v\to 1$, 
but for a different reason: in this limit the HBT-radii go to zero, 
and the whole correlation function becomes entirely flat.

For multiplicity-selected event samples (as they are routinely
collected in heavy ion collisions because multipli\-ci\-ty serves as a
measure for collision centrality) the absence of a minimum in the
correlation function, combined with a clear positive HBT signal at 
$q=0$, can thus be taken as evidence against strong multi-boson 
symmetrization effects and provides a lower limit on the phase-space 
volume (upper limit on the phase-space density) of the source. The 
significance of this limit is controlled by the statistical error on 
the correlation function. The conspicuous absence of a minimum in the 
correlation data known to us supports the evidence for low freeze-out 
phase-space densities (and thus weak multi-boson effects) which was 
obtained with other methods \cite{Fetal}.

On the other hand, searching for such a minimum in event ensembles
with approximately fixed multipli\-ci\-ty appears to be the most direct
approach to find signatures of multi-boson symmetrization effects in 
heavy ion collisions, since it does not rely on model fits  to the
correlation function. It must be noted, however, that 
multiplicity-averaging tends to smear out the minimum, making it
disappear, for instance, in the ``Poisson limit'' of Sec.~\ref{sec3c}. 
To obtain a clear signature thus may require a sharp selection on
multiplicity. In practice the method will be limited by event-by-event 
fluctuations in the slope of the momentum spectra ($\Delta$) and of
the freeze-out radius $R$ which both enter the phase-space volume $v$.
Detailed simulations of such smearing effects go, however, beyond the 
scope of the present paper and will be discussed elsewhere. 

Such a search for multi-boson effects via residual correlations 
requires a 2-Gaussian fit to the correlation function like the one 
obtained by combining according to Eq.~(\ref{62}) the expressions  
(\ref{73}) and (\ref{80}). Compared to the usually employed fit 
functions this doubles the number of HBT radius parameters and thus 
puts very high demands on event statistics. By triggering on central 
collision events in order to exploit azimuthal symmetry of the 
emission region, the number of radius parameters in the single-Gaussian 
fit can in general be reduced to four \cite{reviews} (instead of the 
only two parameters required for the spherically symmetric toy sources 
studied here). For equal-mass collision systems, these can be further 
reduced to two parameters ($R_\perp$ and $R_L$, perpendicular and 
parallel to the beam direction) by selecting pairs with vanishing pair 
momentum in the center-of-mass system \cite{reviews}. To describe the 
residual correlations then requires only two additional parameters, 
$r_\perp$ and $r_L$. The complexity of the resulting fit problem is 
then similar to the now routinely applied 3-dimensional HBT analyses 
in relativistic heavy-ion physics \cite{reviews}, which renders the 
proposed search practicable.

\section{Conclusions}
\label{sec7}

We have studied multi-boson effects on two-particle correlations for
systems with a variety of multiplicity distributions. In contrast to 
the previously studied case of a fixed Poissonian input multiplicity 
distribution at $v\to\infty$, which is then at finite $v$ modified 
by multi-boson effects, we find that in the general case the 
two-particle correlation function is affected not only by 
Bose-Einstein correlations, which can be used to extract source 
size information, but also by residual correlations which depend on 
the pion multiplicity distribution, the average phase-space density 
and the pair momentum.  This is true in particular for events with 
fixed multiplicity.

These residual correlations affect the asymptotic normalization of the 
correlation function at infinite relative momentum and the ``correlation 
strength'', i.e. the ratio of the correlation function at zero and 
infinite relative momentum minus 1. The asymptotic normalization depends 
on the multiplicity distribution, in particular the mean multiplicity 
$\la n\ra$, and the phase-space volume $v^3$. The correlation strength 
depends additionally on the pair momentum $\bK$. Residual correlation 
effects can lead to correlation strengths which are either much larger 
than 1 (for example for the Bose-Einstein multiplicity distribution
at intermediate values of $v$) or much smaller than one and even negative
(for example for fixed multipli\-ci\-ty in the quantum saturation limit 
$v\to 1$). All this happens for completely chaotic sources as those
studied in the present paper and has nothing to do with partial phase
coherence in the particle emission process or other quantum optical 
phenomena. The residual correlations arise, even in the absence of 
HBT correlations, from the generically different weights of pairs, 
triplets, quadruplets etc. of particles {\em within a single event} 
and from {\em mixed events}. The former enter in the numerator, the 
latter in the denominator of the correlation function. These different 
weights play a role when multi-particle symmetrization effects begin to
modify the phase-space distribution of such pairs, triplets, quadruplets
ets. of identical particles, i.e. for sufficiently large phase-space 
densities. The residual correlations vanish only accidentally for very 
specific multiplicity distributions, like those of the ``Poisson limit'' 
in Sec.~\ref{sec3c} or of the ``covariant current formalism'' in Appendix
\ref{app0}.

The separation of residual and genuine Bose-Einstein correlations 
requires a fit of the correlation function with two 3-dimensional 
Gaussians. It thus complicates considerably the usual HBT analysis for
extracting source sizes from two-particle correlation functions. 
Fortunately, we could show that the range of the residual correlations 
and of the Bose-Einstein correlations are defined by well-separated 
length scales such that, at least with sufficiently accurate data, the 
separation is possible. In practice, the demands on the statistical 
accuracy of the data are formidable, and in Section \ref{sec6} we 
therefore suggested to begin such studies by restricting the fits 
to certain kinematic regions (small momenta in the center-of-mass 
system in collisions between equal-mass nuclei) where only a reduced 
number of fit parameters is required.

We have also made a comprehensive model study of multi-boson effects 
on the HBT radii extracted from the Bose-Einstein part of the 
correlation function. In the absence of residual correlations, we showed 
that it is possible to isolate the source size information by first 
dividing out a contamination from ratios of single-particle spectra, 
using measured spectra. In the presence of residual interactions this 
is no longer possible, and the HBT radii cannot be corrected for this
unwanted contamination. In Section \ref{sec5d} we showed how to measure 
it within the model and saw that it becomes dangerously large for 
sources with high phase-space densities, i.e. when multi-boson 
effects become essential.

It is quite likely that the present study eventually turns out to be 
rather academic since Nature only allows particles to decouple from 
the collision region after it has had time to expand to such low 
phase-space densities that multi-boson symmetrization effects become 
negligible. Existing data analyses seem to point in that direction 
\cite{E877_97,Fetal}. However, until we know this for sure we should be 
open to the possibility of strong multi-boson symmetrization phenomena 
and their complicating effect on HBT interferometry. In Section 
\ref{sec6} we have suggested an essentially model-independent way to 
search for such effects which requires little or no prior knowledge 
about the source size (except for that needed to perform final state 
interaction corrections): it is based on the existence of the minimum 
in the (FSI corrected) two-particle correlation function at non-zero 
relative momentum $q$ in events with (approximately) fixed multiplicity. 
If no such minima are found, the results from the present paper can be 
safely shelved, and one may return with strengthened confidence to 
the conventional HBT formalism and source size extraction procedure 
based on Eq.~(\ref{53}).
 
\acknowledgements

The authors thank T.~Cs\"org\H o, S.~Padula, U.A.~Wie\-demann and 
J. Zim\'anyi for stimulating discussions. QHZ was partly supported by  
FAPESP in Sao Paolo, Brazil, the US Department of Energy 
(grant DE-FG03-96ER40972), NSERC of Canada, and the Fonds FCAR of 
the Quebec Government. UH and PS were supported in part by DFG, 
BMBF, and GSI.

\appendix
\section{Relation to the covariant current formalism}
\label{app0}

The general ansatz (\ref{1}), (\ref{7}) for the source density matrix  
on which the present paper is based is formulated in terms of the
unnormalized states (\ref{8}). In this Section we show that the density
matrix of the ``covariant current ensemble'' \cite{GKW,CH94} which 
instead uses normalized states can be brought into a similar form,
with one important difference: in the covariant current ensemble and
in the formalism used in the present paper the normalization of the
density operator and the integration over the phase-space positions at
which the elementary currents are centered are performed {\em in
  opposite order}. We discuss some relevant implications of this
difference.  

The covariant current formalism is based on the density operator 
\cite{GKW,CH94}
 \beq
 \label{01}
   \tilde \rho = \sum_{N=1}^\infty P_N
   \left(\prod_{i=1}^{N}\int d\zeta_i\, \rho(\zeta_i) 
   \int {d\phi_i\over 2\pi} \right) |J\ra \la J| \, ,
 \eeq
where $|J\ra$ is a normalized coherent state,
 \bea
 \label{02}
    |J\ra &\equiv& |J[N;\{\zeta_i,\phi_i\}]\ra
 \nonumber\\
    &=& e^{-\bar n/2}\,\exp\left[ i \int d\tp\,
    \tilde J(\bp)\,\hat a^+_p \right]\, |0\ra\,,
 \eea
with the current $\tilde J(\bp)$ from Eq.~(\ref{4}) and normalization
$\exp(-\bar n/2)$ with $\bar n$ given in (\ref{10}). It corresponds to 
a specific pion multiplicity distribution $\tilde p_n$ which can be 
calculated by evaluating
 \bea
 \label{03}
   \tilde p_n &=& {\la n | \hat\rho| n\ra \over \la n|n\ra }
 \\
   &=& \sum_{N=1}^\infty P_N
   \left(\prod_{i=1}^{N}\int d\zeta_i\, \rho(\zeta_i) 
   \int {d\phi_i\over 2\pi} \right) {\ |\la n|J\ra|^2\over \bar n^n}
 \nonumber
 \eea
with the states (\ref{8}). Since
 \beq
 \label{04}
   \la n|J\ra = e^{-\bar n/2} {\bar n^n \over \sqrt{n!}}\,,
 \eeq 
each coherent state itself has a Poissonian multiplicity distribution:
 \beq
 \label{05}
   {|\la n|J\ra|^2\over \la n|n\ra} = {\bar n^n\over n!}\, e^{-\bar n}\,.
 \eeq
Thus we have
 \beq
 \label{06}
   \tilde p_n = \sum_{N=1}^\infty P_N
   \left(\prod_{i=1}^{N}\int d\zeta_i\, \rho(\zeta_i) 
   \int {d\phi_i\over 2\pi} \right) {\bar n^n\over n!}\,e^{-\bar n}
   \, ,
 \eeq
where $\bar n$ depends on all integration variables. By summing over $n$
and using $\sum_n e^{-\bar n} (\bar n^n/n!)=1$ and the nor\-ma\-lization
of $P_N$ as well as (\ref{2}), one easily checks that $\tilde p_n$ is
normalized: 
 \beq
 \label{07}
   \sum_{n=0}^\infty \tilde p_n =1\,.
 \eeq

We can decompose the coherent state projection operator in (\ref{01}) 
into the set (\ref{8}) of unnormalized $n$-particle states $|n\ra$:
 \bea
 \label{08}
   |J\ra \la J| &=& \sum_{n,m=0}^\infty 
   {|n\ra\la n|J\ra\la J|m\ra\la m|\over \la n|n\ra \la m|m\ra}
 \nonumber\\
   &=& \sum_{n,m=0}^\infty {e^{-\bar n} \over 
                          \sqrt{n!\,m!}} |n\ra\la m| \,.
 \eea
For the calculation of the $n$-particle spectra, which always involve
equal numbers of creation and annihilation operators, only the
diagonal terms $n=m$ contribute. We can thus replace the density
operator $\tilde\rho$ of the covariant current ensemble (\ref{01}) by 
 \beq
 \label{09}
   \tilde\rho' = 
   \sum_{N=1}^\infty P_N 
   \left(\prod_{i=1}^{N}\int d\zeta_i\, \rho(\zeta_i) 
   \int {d\phi_i\over 2\pi} \right) {e^{-\bar n}\over n!} 
   |n\ra\la n|\,.
 \eeq
Writing this in the form (\ref{1}), namely
 \beq
 \label{010}
   \tilde\rho' = \sum_{n=0}^\infty \tilde p_n\,
   \tilde\rho_n\,,
 \eeq
one finds for the density operator of the subspace corresponding to
events with fixed particle multiplicity $n$ the expression
 \beq
 \label{011}  
   \tilde\rho_n = {1\over n!\,\tilde p_n}
   \sum_{N=1}^\infty P_N
   \left(\prod_{i=1}^{N}\int d\zeta_i\, \rho(\zeta_i) 
   \int {d\phi_i\over 2\pi} \right)
    e^{-\bar n}\, |n\ra \la n|\, .
 \eeq
When comparing this with the ansatz (\ref{7}) a characteristic 
difference becomes apparent: the covariant current formalism involves
an additional weight function $\exp(-\bar n[N;\{\zeta_i,\phi_i\}])$
under the integral over the positions and phases of the elementary
source currents $j_0$. This weight results from the use of normalized
coherent states.

An arbitrary weight function would at this point stop further analytic 
progress. For the specific weight function associated with the use of
normalized coherent states, however, all $n$-particle spectra can
still be calculated ana\-lytically \cite{GKW,CH94}, in fact with much 
less effort than for the ansatz (\ref{7}). This is due to the
simplifications resulting from the fact that coherent states are
eigenstates of the annihilation operator. For example, the single
particle spectrum is given by
 \bea
 \label{012}
    &&N_1(\bp) = E_p\,\tr\bigl(\tilde\rho'\,\hat a^+_p \hat a_p\bigr)
 \\
    &&= E_p \sum_{N=1}^\infty P_N
   \left(\prod_{i=1}^{N}\int d\zeta_i\, \rho(\zeta_i) 
   \int {d\phi_i\over 2\pi} \right)
   |\tilde J(\bp)|^2\,\la J|J\ra\, .
 \nonumber
 \eea
We write
 \beq
 \label{013}
   \la J|J\ra = 1 = \sum_{n=0}^\infty {\bar n^n\over n!}\, e^{-\bar n}
 \eeq
and calculate the normalization of the single particle spectrum as
 \bea
 \label{014}
   &&\int d\tp\,N_1(\bp) 
 \nonumber\\
   &&= \sum_{N=1}^\infty P_N
   \left(\prod_{i=1}^{N}\int d\zeta_i\, \rho(\zeta_i) 
   \int {d\phi_i\over 2\pi} \right)
   \bar n \sum_{n=0}^\infty {\bar n^n\over n!}\, e^{-\bar n}
 \nonumber\\
    &&= \sum_{N=1}^\infty P_N
   \left(\prod_{i=1}^{N}\int d\zeta_i\, \rho(\zeta_i) 
   \int {d\phi_i\over 2\pi} \right)
   \sum_{m=0}^\infty m\,{\bar n^m\over m!}\, e^{-\bar n}
 \nonumber\\
    &&= \sum_{m=0}^\infty m\, \tilde p_m = \la m \ra\,.
 \eea
The single particle spectrum is thus correctly norma\-lized to the
average particle multiplicity, calculated from the multiplicity
distribution $\tilde p_n$ of the covariant current ensemble.
Similarly one finds that the two-particle spectrum
 \bea
 \label{015}
    N_2(\bp_1,\bp_2) &=& 
    E_{p_1} E_{p_2}\sum_{N=1}^\infty P_N
   \left(\prod_{i=1}^{N}\int d\zeta_i\, \rho(\zeta_i) 
   \int {d\phi_i\over 2\pi} \right)
 \nonumber\\
   &&\times\,|\tilde J(\bp_1)|^2\,|\tilde J(\bp_2)|^2\,\la J|J\ra
 \eea
is correctly normalized to twice the average number of pairs
calculated from $\tilde p_n$:
 \beq
 \label{016}
   \int d\tp_1\,d\tp_2\,N_2(\bp_1,\bp_2) = 
   \sum_{m=0}^\infty m(m{-}1)\, \tilde p_m = \la m(m{-}1) \ra\,.
 \eeq
From the studies in \cite{GKW,CH94} it is known that the
norma\-li\-zation (\ref{014}) can be alternatively written as
 \beq
 \label{017}
   \la m \ra = \lda N \rda \, n_0\,,
 \eeq
where $\lda N \rda$ is the average number of elementary currents and
$n_0$ is their normalization (\ref{6}). The two-particle correlation
function is in the covariant current formalism given by 
 \beq
 \label{018}
   C(\bq,\bK) = {\lda N(N{-}1) \rda\over \lda N\rda^2} 
   \left( 1 + 
   {\left\vert\int_x S(x,\bK)\,e^{iq\cdot x}\right\vert^2 \over
              \int_x S(x,\bp_1) \int_y S(x,\bp_2)}\right)
 \eeq
where the effective emission function is nothing but the Wigner density
associated with the currents $J$:
 \beq
 \label{019}
   S(x,\bK) = \int\frac{d^4y}{2(2\pi)^3}\, e^{-iK{\cdot}y}
   \left\langle J^*(x+\half y)J(x-\half y)\right\rangle \,.
 \eeq   
This is, up to the prefactor, the standard form (\ref{53}). The
covariant current ensemble with its multiplicity distribution $\tilde
p_n$ is thus a second example (besides the ``Poisson limit'' of
Section~\ref{sec3c}) for a system in which multi-boson effects do not
cause residual correlations. Up to the different normalization ($\lda
N(N{-}1)\rda$ vs. 1), the only difference between the density
operator (\ref{7}) in the ``Poisson limit'' and the covariant current
ensemble with the density operator (\ref{09}) is the different
definition of the effective emission function: in the former case it
is given by Eqs.~(\ref{51}) and (\ref{52}) while for the latter the
much simpler definition (\ref{019}), with all multi-boson effects
already fully accounted for. The price to be paid is a more
complicated expression for the multiplicity distribution ((\ref{06})
vs. (\ref{40}) -- while, for Gaussian sources, the latter is known
analytically, the former seems to require numerical evaluation). In
both cases the measured multiplicity distribution ($\bar p_n$ and
$\tilde p_n$, respectively) depends in a complicated way on the mean
phase-space density $d$ of the source.

\section{Calculating the correlation functions}
\label{appa}

In this Section we give some technical steps for the analytical evaluation 
of the correlation functions for the Gaussian sources defined in 
Sec.~\ref{sec4}.

Let us define
 \begin{mathletters}
 \label{A1}
 \bea 
 \label{A1a}
   A_{ij}(\kappa) &=& h_{i{+}j}\, c_i\, c_j\,
   e^{-{\kappa^2\over 2}\bigl({1\over a_i}+{1\over a_j}\bigr)}\, ,
 \\
 \label{A1b}
   B_{ij}(\kappa) &=& h_i\, h_j\, c_i\, c_j\, 
   e^{-{\kappa^2\over 2}\bigl({1\over a_i}+{1\over a_j}\bigr)}\, ,
 \\
 \label{A1c}
   A(\kappa) &=& \sum_{i,j=1}^\infty A_{ij}(\kappa)\, ,
 \\
 \label{A1d}
   B(\kappa) &=& \sum_{i,j=1}^\infty B_{ij}(\kappa)\,,
 \eea
 \end{mathletters}
as well as
 \beq
 \label{A2}
   E_{ij}(\tilde\bq,\bbox{\kappa}) =
   e^{-{\tilde q^2\over 2 v^2}\bigl({1\over a_i}{+}{1\over a_j}\bigr)}
   \cosh\left[{\tilde\bq{\cdot}\bbox{\kappa}\over v}
   \left({1\over a_i}{-}{1\over a_j}\right)\right].
 \eeq
A few simple steps, using eqs.~(\ref{57}) and (\ref{63}) and exploiting 
the symmetry of the sums under exchange of $i$ and $j$, lead from
(\ref{62}) to
 \bea
 \label{A3}
   C_2^{({\rm res})}(\tilde\bq,\bbox{\kappa}) &=&
   {\sum_{i,j=1}^\infty A_{ij}(\kappa) \, E_{ij}(\tilde\bq,\bbox{\kappa})
    \over
    \sum_{i,j=1}^\infty B_{ij}(\kappa) \, E_{ij}(\tilde\bq,\bbox{\kappa})}
   \,,
 \\
 \label{A4}
   R_2(\tilde\bq,\bbox{\kappa}) &=&
   {\sum_{i,j=1}^\infty A_{ij}(\kappa) \, 
                        e^{-{\tilde q^2\over 2}(a_i+a_j)}
    \over
    \sum_{i,j=1}^\infty A_{ij}(\kappa) \, E_{ij}(\tilde\bq,\bbox{\kappa})}
   \,.
 \eea

For very dilute ($v\to\infty$) and very dense ($v\to 1$) sources
Eqs.~(\ref{60}) and (\ref{61}) can be used to calculate analytically
the behaviour of the correlation function in the limits $q\to 0$ and 
$q\to\infty$. Note that in these limits the dependence on the angle 
between $\tilde\bq$ and $\bbox{\kappa}$ (which enters via (\ref{A2}))
disappears such that all following results depend only on the modulus
$\kappa$ of the pair momentum:

\medskip
\noindent $\bbox{1.\ q\to\infty:}$
\medskip

\noindent For $v\to\infty$ the $q^2$-dependent exponential terms in the
numerator of $R_2$ are much smaller than those in the denominator, and
the Bose-Einstein correlations thus va\-nish as $q\to\infty$. For $v\to 1$
the analysis is only slightly harder: writing $v=1+\epsilon$, one finds 
that the exponential terms of both the numerator and denominator are given 
by $\exp[-\tilde q^2(1-\epsilon)]$ for all values of $i,j$ except
for the lowest order term with $i=j=1$; for the latter the numerator 
goes as $\exp[-\tilde q^2]$ while the denominator goes as 
$\exp[-\tilde q^2(1-2\epsilon)]$. Again the numerator vanishes 
more quickly, and the Bose-Einstein correlations disappear at $q\to\infty$,
as they should.

For the residual correlations (\ref{A3}) one checks similarly that in 
both limits ($v\to\infty$ and $v\to 1$) for $q\to\infty$ numerator and 
denominator are both dominated by the term $i=j=1$, giving
 \beq
 \label{A5}
   \lim_{q\to\infty} C_2^{({\rm res})}(\tilde\bq,\bbox{\kappa}) =
   {A_{11}\over B_{11}} = {h_2\over h_1^2} \equiv {\cal N}\,.
 \eeq

\medskip
\noindent $\bbox{2.\ q\to 0:}$
\medskip

\noindent In this limit $R_2$ approaches trivially the value 1 (see
(\ref{62})), while the residual correlations are given by
 \beq
 \label{A6}
   \lim_{q\to 0} C_2^{({\rm res})}(\tilde\bq,\bbox{\kappa})
   = {A(\kappa)\over B(\kappa)}\,.
 \eeq
The incoherence parameter defined in Eq.~(\ref{incoh}) is thus 
$\lambda_{\rm incoh}(\kappa)=1$, whereas the correlation strength 
(\ref{33}) is given by (\ref{69}).

\medskip

Other useful limiting expressions are
 \bea
 \label{A7}
   &&\lim_{v\to\infty} h_1 = \la  n\ra\,,
 \nonumber\\
   &&\lim_{v\to\infty} h_2 = \la  n(n{-}1)\ra\,,
 \nonumber\\
   &&\lim_{v\to\infty} h_i = \la  n(n{-}1)\cdots(n{-}i{+}1)\ra
   \ \ {\rm for}\ i\geq 2\,,
 \\
 \label{A8}
   &&\lim_{v\to\infty} A(\kappa) = \lim_{v\to\infty} A_{11}(\kappa) = 
   \la  n(n{-}1)\ra\, e^{-\kappa^2}\,,
 \\
 \label{A9}
   &&\lim_{v\to\infty} B(\kappa) = \lim_{v\to\infty} B_{11}(\kappa) = 
   \la  n\ra^2 \, e^{-\kappa^2}\,,
 \eea
and (to linear accuracy in $\epsilon$)
 \bea
 \label{A10}
   &&\lim_{v\to 1+\epsilon} h_i = 
     \bigl(1+{\textstyle{3\over2}}i\epsilon\bigr)
     \sum_{n=i}^\infty p_n - {\textstyle{3\over2}} \epsilon p_i
     \ \ {\rm for}\ i\geq 1\, ,
 \\
 \label{A11}
   &&\lim_{v\to 1+\epsilon} A(\kappa) = {\la  n(n-1)\ra\over 2}
   e^{-\kappa^2} 
 \\
   &&\qquad\times 
   \left[ 1 - (\kappa^2{-}3) \epsilon 
   + {\la  n\ra - (1{-}p_0) \over \la  n(n{-}1)\ra}
     (2\kappa^2{-}3)\epsilon \right]\,,
 \nonumber\\
 \label{A12}
   &&\lim_{v\to 1+\epsilon} B(\kappa) = \la  n\ra^2 e^{-\kappa^2}
   \left[ 1 - {\la  n\ra - (1{-}p_0) \over 
               \la  n\ra}(\kappa^2{-}3) \epsilon \right]\,.
 \nonumber\\
 \eea
These last identities require the evaluation of the following sums involving
$\tilde h_i \equiv \sum_{n=i}^\infty p_n$:
 \bea
 \label{A13}
   &&\sum_{i=2}^\infty \tilde h_i = \la  n\ra - (1{-}p_0)\,,
 \\
 \label{A14}
   &&\sum_{i,j=1}^\infty \tilde h_{i{+}j} = 
     {\la  n(n{-}1)\ra\over 2}\,,
 \\
 \label{A15}
   &&\sum_{i,j=1}^\infty \tilde h_{i{+}j{+}1} = 
     {\la  (n{-}1)(n{-}2)\ra\over 2} - p_0\,,
 \eea
which is achieved by appropriate reordering and relabelling of the nested 
sums.

\section{Calculating the radius parameters}
\label{appb}

Inserting (\ref{A4}) into (\ref{75}) and separating the constant
and $\theta$-dependent terms one finds
 \bea
 \label{B1}
   &&{R_\perp^2(\kappa)\over R^2} = 
   \sum_{i,j=1}^\infty
   {A_{ij}(\kappa)\over 2\,A(\kappa)}  \Bigl[ (a_i{+}a_j) 
   - {1\over v^2}\Bigl({1\over a_i} + {1\over a_j}\Bigr)\Bigr]\,,
 \\
 \label{B2}
   && {R_\parallel^2(\kappa)-R_\perp^2(\kappa)\over R^2} =
   {\kappa^2\over v^2} \sum_{i,j=1}^\infty
   {A_{ij}(\kappa)\over 2\,A(\kappa)}  
   \Bigl({1\over a_i} - {1\over a_j}\Bigr)^2\,.
 \eea
In the dilute gas limit $v\to\infty$ only the lowest order terms 
$i{=}j{=}1$ contribute to the sums, and one finds the expressions
(\ref{76}). In the opposite limit $v\to 1$ one can use 
(\ref{A10})--(\ref{A12}) to obtain
 \begin{mathletters}
 \label{B3}
 \bea
 \label{B3a}
   \lim_{v\to 1} r_1 &=& 
   2 {\la  n\ra - (1{-}p_0) \over \la  n(n-1)\ra}
 \\ 
   &=& \left\{ \begin{array}{ll}
              {2\over n} &\ \ {\rm for\ fixed}\ n\geq 2\,,\\
              {1\over 1+\la  n\ra} &\ \ {\rm Bose{-}Einstein\,,}\\
              2{\la  n\ra - 1 + e^{-\la  n \ra}
              \over \la  n\ra^2} &\ \ {\rm Poisson\,,}
             \end{array}\right.
 \nonumber\\
 \label{B3b}
   \lim_{v\to 1} r_2 &=& 0\,.
 \eea
 \end{mathletters}
All these limits are $\kappa$-independent; however, in the intermediate 
region $1<v<\infty$ the HBT radii develop a $\kappa$-dependence as 
shown by the numerical results in Sec.~\ref{sec5d}. 

The vanishing of $r_2$ in the limit $v\to 1$ has an interesting origin: 
in this limit the functions $G_i(\bp_1,\bp_2)$ in (\ref{57}) become all 
identical, and the single particle spectrum (\ref{36}) is again a simple 
Gaussian with width $\Delta$, just as in the opposite limit $v\to\infty$. 
In this case the angle-dependent term $\sim\bq{\cdot}\bK$ drops out from 
the product of single particle spectra in the denominator of the 
Bose-Einstein correlator $R_2$, and the two HBT radii $R_\perp$ and 
$R_\parallel$ become equal.  

The ``residual correlation radii'' are obtained by inserting (\ref{A3})
into (\ref{81}) and separating the constant and $\theta$-dependent pieces:
 \bea
 \label{B4}
   &&{r_\perp^2(\kappa)\over R^2} = 
   {A(\kappa)\over {\cal N}\,B(\kappa)-A(\kappa)}\, 
 \\
   &&\quad \times
   {1\over 2v^2} \sum_{i,j=1}^\infty 
   \Bigl({B_{ij}(\kappa)\over B(\kappa)}-{A_{ij}(\kappa)\over A(\kappa)}\Bigr) 
   \Bigl({1\over a_i} + {1\over a_j}\Bigr)\,,
 \nonumber\\
 \label{B5}
   &&{r_\parallel^2(\kappa)-r_\perp^2(\kappa)\over R^2} =
   {A(\kappa)\over {\cal N}\,B(\kappa)-A(\kappa)}
 \\
   &&\quad \times   
   {\kappa^2\over 2v^2}\, \sum_{i,j=1}^\infty 
   \Bigl({A_{ij}(\kappa)\over A(\kappa)}-{B_{ij}(\kappa)\over B(\kappa)}\Bigr)
   \Bigl({1\over a_i} - {1\over a_j}\Bigr)^2\,.
 \nonumber
 \eea
To facilitate the discussion in the main text (Section \ref{sec5e}) we
here give a specific expression in the limit $v\to 1$. We decompose the
r.h.s. of (\ref{B4}) into three factors which we expand in powers of 
$\epsilon = v-1$:
 \begin{mathletters}
 \label{B6} 
 \bea
 \label{B6a}
   &&\lim_{v\to 1+\epsilon} A(\kappa) = a_0 + a_1\,\epsilon +\dots\,,
 \\
 \label{B6b}
   &&\lim_{v\to 1+\epsilon} \bigl( {\cal N}\,B(\kappa) - A(\kappa) \bigr) 
     = b_0 + b_1\,\epsilon +\dots\,,
 \\
 \label{B6c}
   &&\lim_{v\to 1+\epsilon} {1\over 2v^2}\sum_{ij}\dots =
     c_0 + c_1\,\epsilon +\dots\,.
 \eea
 \end{mathletters}
The lowest order terms have been calculated (see also (\ref{A11})):
 \begin{mathletters}
 \label{B7} 
 \bea
 \label{B7a}
   a_0 &=& {\la  n(n{-}1)\ra\over 2} e^{-\kappa^2}\,,
 \\ 
 \label{B7b}
   b_0 &=& \left( {1{-}p_0{-}p_1\over (1{-}p_0)^2} \la  n\ra^2
   -{\la  n(n{-}1)\ra\over 2}\right) e^{-\kappa^2}\,,
 \\ 
 \label{B7c}
   c_0 &=& 0\,,
 \\ 
 \label{B7d}
   c_1 &=& - {1{-}p_0\over\la  n\ra} 
   + 2 {\la  n(n{-}1)\ra\over 2\la  n\ra^2}\,.
 \eea
 \end{mathletters}
Using Eq.~(\ref{B3a}) in the form
 \beq
 \label{B8}
  \lim_{v\to 1+\epsilon} R_\perp^2(\kappa) = 
  2 {\la  n\ra - (1{-}p_0) \over \la  n(n{-}1)\ra}\, 
  2\,\epsilon\,R^2
 \eeq
one finds
 \beq
 \label{B9}
   \lim_{v\to 1} {r_\perp^2(\kappa)\over R_\perp^2(\kappa)}
   = {1\over 4} {\la  n(n{-}1)\ra \over \la n\ra - (1{-}p_0)}
     {a_0\, c_1\over b_0} \,.
 \eeq
This relation is the basis of the discussion in Sec.~\ref{sec5e}. 

For the Bose-Einstein multiplicity distribution (\ref{64d}) an unfortunate 
coincidence causes both $b_0$ and $c_1$ to va\-nish. In this case their 
ratio in (\ref{B9}) must be replaced by $c_2/b_1$. The calculation of 
$c_2$ requires a consistent calculation to order $\epsilon^2$ which we
have not done. Therefore, for the case of a Bose-Einstein distribution 
we have no analytical expression for the limit (\ref{B9}).



\begin{thebibliography}{100}
\bibitem{fn1}
  In \cite{Geiger}, Eq.~(38), an incorrect expression was given for 
  $R_s^2$ which resulted from an incorrect rewriting of the correct 
  expression given in \cite{Cris98_2}. Since the numerical simulations
  in \cite{Geiger} were compared with the original correct analytical
  expression, all figures in \cite{Geiger} are correct, although the
  discussion in the paragraph following Eq.~(38) in \cite{Geiger} is
  inaccurate and slightly misleading.
\end{thebibliography}
\end {document}